# High-frequency measurements of aeolian saltation flux: Field-based methodology and applications

Raleigh L. Martin[a*], Jasper F. Kok[a], Chris H. Hugenholtz[b], Thomas E. Barchyn[b], Marcelo Chamecki[a], Jean T. Ellis[c]

[a]Department of Atmospheric and Oceanic Sciences, University of California, Los Angeles, CA 90095, USA.

[b]Department of Geography, University of Calgary, Calgary, AB, Canada.

[c]Department of Geography and School of Earth, Ocean and Environment, University of South Carolina, Columbia, SC 29208, USA

[*]**Corresponding Author**
Raleigh L. Martin
Department of Atmospheric and Oceanic Sciences
University of California, Los Angeles
520 Portola Plaza
Los Angeles, CA 90095
e-mail: raleighm@atmos.ucla.edu

## Abstract
Aeolian transport of sand and dust is driven by turbulent winds that fluctuate over a broad range of temporal and spatial scales. However, commonly used aeolian transport models do not explicitly account for such fluctuations, likely contributing to substantial discrepancies between models and measurements. Underlying this problem is the absence of accurate sand flux measurements at the short time scales at which wind speed fluctuates. Here, we draw on extensive field measurements of aeolian saltation to develop a methodology for generating high-frequency (25 Hz) time series of total (vertically-integrated) saltation flux, namely by calibrating high-frequency (HF) particle counts to low-frequency (LF) flux measurements. The methodology follows four steps: (1) fit exponential curves to vertical profiles of saltation flux from LF saltation traps, (2) determine empirical calibration factors through comparison of LF exponential fits to HF number counts over concurrent time intervals, (3) apply these calibration factors to subsamples of the saltation count time series to obtain HF height-specific saltation fluxes, and (4) aggregate the calibrated HF height-specific saltation fluxes into estimates of total saltation fluxes. When coupled to high-frequency measurements of wind velocity, this methodology offers new opportunities for understanding how aeolian saltation dynamics respond to variability in driving winds over time scales from tens of milliseconds to days.

## 1. Introduction
Wind-blown (aeolian) transport of sand shapes a variety of desert, coastal, and planetary landscapes (e.g., Lancaster, 1988; Bridges et al., 2012; Durán and Moore, 2013). Saltation, the ballistic hopping motion of wind-blown sand grains, drives the bulk of aeolian sand flux (Bagnold, 1941), abrades bedrock (Perkins et al., 2015), erodes soil (Chepil, 1945), and generates airborne dust through impacts with the soil surface (Gillette et al., 1974; Shao et al., 1993; Marticorena and Bergametti, 1995; Kok et al., 2014). Studies of landscape evolution and dust generation require models that accurately relate wind speed, surface conditions, and the resulting sand flux (e.g., Kok et al., 2012).

Unfortunately, aeolian saltation models often do a poor job of predicting rates of sand transport in natural environments (e.g., Kok et al., 2012; Sherman and Li, 2012; Sherman et al., 2013; Barchyn et al., 2014b). Most existing aeolian saltation models are based on the assumption of a steady-state momentum balance (e.g., Ungar and Haff, 1987; Andreotti, 2004), but saltation in natural environments is driven by widely-varying turbulence spectra, which produce broad spatial and temporal variations in the saltation flux (e.g., Baas, 2006; Durán et al., 2011). Though wind-tunnel experiments can capture some of this turbulent variability (e.g., Li and McKenna Neuman, 2012, 2014), such experimental settings differ substantially from natural environments in their ability to capture saltation-wind interactions (e.g., Sherman and Farrell, 2008), producing broad unexplained discrepancies between field and laboratory measurements (Barchyn et al., 2014b; Martin and Kok, 2017a). In addition, sedimentological factors like soil moisture (Arens, 1996; Davidson-Arnott et al., 2008), surface crusting (Gillette et al., 1982), sediment availability (Webb et al., 2016a), electrification (e.g., Kok and Renno, 2008), mid-air particle collisions (e.g., Sørensen and McEwan, 1996; Carneiro et al., 2013), and surface grain-size distributions (Iversen and Rasmussen, 1999), cause further differences among models, wind tunnel experiments, and field measurements. Though recent studies have sought to understand each individual factor governing aeolian saltation dynamics separately, our ability to model how this constellation of atmospheric and sedimentological factors together controls the aeolian saltation flux remains limited. Coupled high-frequency (HF) field measurements of wind and saltation offer the potential to improve our understanding of some of the atmospheric factors affecting saltation flux variability, and they can also help to constrain the role of sedimentological factors (e.g., Martin and Kok, 2017a).

Until recently, however, field-based observations of aeolian saltation have been limited to low-frequency (LF) measurements with saltation traps (i.e., sampling interval ≥ ~1 minute). These traps generally provide reliable measures of the saltation mass flux over time scales of minutes to days (e.g., Greeley et al., 1996; Sherman et al., 1998; Namikas, 2003), though comparative studies reveal variations in trap accuracy that depend on wind speed and airborne grain sizes (e.g., Goossens et al., 2000). Assuming that saltation traps do indeed provide reliable measures of the saltation flux, LF studies are useful for relating saltation flux and vertical saltation profile characteristics (Greeley et al., 1996; Namikas, 2003; Farrell et al., 2012) to time-averaged wind speeds (Sherman et al., 1998; Sherman and Li, 2012). However, such LF studies are unable to resolve the HF spatial and temporal variability in saltation flux (i.e., ≤~1 minute) resulting from wind turbulence in the atmospheric boundary layer (e.g., Baas and Sherman, 2005). Furthermore, estimates of total (vertically-integrated) saltation flux typically depend on fitting curves to vertical profiles of saltation flux, but there remains disagreement over fitting protocols (Ellis et al., 2009a) and over the proper functional form for these fits: exponential (Dong et al., 2012; Fryrear and Saleh, 1993; Namikas, 2003), power law (Zobeck and Fryrear, 1986), or some hybrid of these (Bauer and Davidson-Arnott, 2014; Dong et al., 2011). Such variability and fitting uncertainty is thought to produce much of the disagreement between measurements and models of aeolian saltation flux (Barchyn et al., 2014b).

To better resolve turbulence-induced saltation fluctuations, a variety of new HF sensors have been deployed in field studies over the past two decades (e.g., Baas, 2004; Barchyn and Hugenholtz, 2010; Sherman et al., 2011). HF measurements typically register individual sand grains, using sensors with optical gates (e.g., Hugenholtz and Barchyn, 2011; Etyemezian et al., 2017), piezoelectric impact plates (e.g., Barchyn and Hugenholtz, 2010; Sherman et al., 2011), or acoustic microphones (e.g., Spaan and van den Abeele, 1991; Ellis et al., 2009b). Such sensors are capable of recording measurements at time scales ranging from 10s of milliseconds to seconds (e.g., Sterk et al., 1998; Baas, 2008; Martin et al., 2013), much faster than the most rapid automated saltation trap sampling techniques (Bauer and Namikas, 1998; Butterfield, 1991; Namikas, 2002; Ridge et al., 2011).

HF saltation sensors have been used in recent years to address a variety of questions in aeolian research. Recent field studies deploying HF saltation sensors, in tandem with HF anemometer wind observations, have quantified the frequently observed spatial and temporal patterns of alternating high and low saltation flux known as "aeolian streamers" (Baas and Sherman, 2005; Weaver and Wiggs, 2011). Other HF field deployments have offered further insight on the temporal variability of saltation flux (Sterk et al., 1998; Schonfeldt and von Lowis, 2003; Baas, 2006; Martin et al., 2013), across complex topography (Bauer et al., 2012, 2015; Davidson-Arnott et al., 2012; Hoonhout and de Vries, 2017) and around vegetation (Barrineau and Ellis, 2013; Chapman et al., 2013). HF sensors are also vital for describing saltation intermittency and thresholds (Stout and Zobeck, 1997; Schönfeldt, 2004; Wiggs et al., 2004a; Barchyn and Hugenholtz, 2011; Poortinga et al., 2015; Webb et al., 2016a) and the effects of humidity and soil moisture on these thresholds (Arens, 1996; Wiggs et al., 2004b; Delgado-Fernandez et al., 2012). Optical HF sensors have also been used extensively for the measurement of wind-driven snow transport (Leonard et al., 2011; Bellot et al., 2013; Maggioni et al., 2013; Trujillo et al., 2016).

Though helpful for advancing the understanding of saltation dynamics, HF measurements typically provide only relative, not absolute, measures of the aeolian saltation flux (e.g., Barchyn et al., 2014a). Typically, these HF sensors produce data in counts per second. Such count rates are only internally relative and require a conversion to physically meaningful quantities, which may not be as simple as one grain per count (Barchyn et al., 2014a). For certain research purposes, these relative HF saltation measurements are sufficient, such as for studies of saltation intermittency and thresholds at a single point (e.g., Stout and Zobeck, 1997; Martin and Kok, 2017b). However, to understand the relationship between turbulence structures and saltation flux variability in space and time, absolute HF measurements of saltation flux are needed (e.g., Martin et al., 2013; Bauer et al., 2015; Hoonhout and de Vries, 2017).

To address the need for reliable HF saltation flux measurements, studies have compared the performance of different HF sensors (e.g., Davidson-Arnott et al., 2009; Leonard et al., 2011; Massey, 2013) and assessed the comparability of HF particle counts to LF trap saltation fluxes (e.g., Sterk et al., 1998; Goossens et al., 2000; Sherman et al., 2011). Though these studies generally reveal linear relationships among particle counts from different sensors (e.g., Barchyn et al., 2014a), they also indicate substantial differences in sensitivity between sensors of the same type (Baas, 2008) or among sensors of different types (Hugenholtz and Barchyn, 2011). HF saltation sensors are potentially subject to "saturation" effects – i.e., reaching a maximum saltation flux above which measured particle counts no longer increase (Hugenholtz and Barchyn, 2011; Sherman et al., 2011). HF sensors may also have response sensitivities to momentum or particle size (Barchyn et al., 2014a). Additionally, HF sensors may display "drift", or variation in their performance through time, due to environmental conditions causing sensor degradation (Hugenholtz and Barchyn, 2011; Bauer et al., 2012; Barchyn et al., 2014a).

A fundamental issue with most HF measurements is that, whereas traps and sensors typically provide only height-specific values for the saltation flux, models of aeolian saltation consider total (vertically-integrated) saltation fluxes (e.g., Bagnold, 1941; Owen, 1964; Ungar and Haff, 1987; Andreotti, 2004). Therefore, it is necessary to calculate vertical profile fits in order to estimate the total saltation flux. In addition to issues of deciding the best form for such profile fits (described above), turbulent variability and counting uncertainties may hinder the convergence of vertical flux profiles over short time scales (e.g., Bauer and Davidson-Arnott, 2014). Thus, existing studies of high-frequency saltation flux variability are limited to examination of relative or height-specific saltation fluxes (e.g., Baas, 2008).

In short, LF trap and HF sensor measurement techniques each have distinctive advantages and disadvantages for determining saltation flux. LF measurements can accurately measure horizontal and vertical profiles of mass flux and sediment size, but they can detect only the broadest fluctuations of saltation mass flux associated with the passage of large-scale turbulent structures (McKenna Neuman et al., 2000). HF sensors can resolve saltation responses to turbulence, but their ability to provide absolute mass fluxes is questionable (Hugenholtz and Barchyn, 2011). Ideally, the respective advantages of LF and HF measurements could be combined to provide HF time series of absolute saltation flux.

In this paper, we describe a new methodology to generate reliable high-resolution time series of the total (vertically-integrated) saltation mass flux. Specifically, we do so by combining LF measurements from sediment traps with HF measurements from optical particle counters. As the basis for explaining this methodology, we first describe the three field sites at which we collected LF and HF saltation measurements (Section 2), and we describe the instrumentation involved (Section 3). In Section 4 we describe the sequence of steps for obtaining calibrated, high-frequency measurements of the total saltation flux, and we present some illustrative results. In the Discussion (Section 5), we outline the advances and limitations of the HF saltation calibration methodology, and we offer guidelines for future field deployments. We conclude and summarize our findings in Section 6.

## 2. Field sites

In this section, we provide basic information about the three field sites for our data collection and the grain-size characteristics of the surface sediments at these sites.

**Table 1. Summary information for field sites**

|  | **Jericoacoara** | **Rancho Guadalupe** | **Oceano** |
|---|---|---|---|
| Coordinates | 2.7969°S, 40.4837°W | 34.9592°N, 120.6431°W | 35.0287°N, 120.6277°W |
| Deployment dates | 2014-11-13 to 2014-11-20 | 2015-03-23 to 2015-03-24 | 2015-05-15 to 2015-06-04 |
| Distance to shoreline (m) | 750 | 950 | 650 |
| Sand patch distance (m) | 250 | 100 | 300 |
| Anemometer heights, $z_U$ (m) | *2014-11-13:* 0.48, 0.97, 1.93, 3.02 *2014-11-14*: 0.47, 0.98, 1.76, 2.59 *2015-11-20:* 0.47, 0.97, 1.76, 2.58 | *2015-03-23:* 0.45, 1.00, 1.80, 2.75 *2015-03-24:* 0.43, 1.00, 1.79, 2.81 | *All dates:* 0.64, 1.16, 2.07, 3.05, 6.00, 8.95 |
| LF trap (BSNE) heights, $z_{LF,i}$ (m) | 0.10 – 0.52 | 0.08 – 0.70 | 0.05 – 0.46 |
| HF sensor (Wenglor) heights, $z_{HF,i}$ (m) | 0.02 – 0.29 | 0.02 – 0.32 | 0.06 – 0.47 |
| 10th percentile grain diameter, $d_{10}$ (mm) | 0.097 ± 0.012 | 0.219 ± 0.035 | 0.190 ± 0.032 |
| median grain diameter, $d_{50}$ (mm) | 0.526 ± 0.037 | 0.533 ± 0.026 | 0.398 ± 0.070 |
| 90th percentile grain | 0.847 ± 0.037 | 0.839 ± 0.034 | 0.650 ± 0.075 |

| diameter, $d_{90}$ (mm) | | | |
| Active saltation days | 3 | 2 | 12 |

## 2.1. General information about field sites

We measured aeolian saltation at three deployment sites: Jericoacoara, Ceará, Brazil (2.7969°S, 40.4837°W); Rancho Guadalupe, California, United States (34.9592°N, 120.6431°W); and Oceano, California, United States (35.0287°N, 120.6277°W). The Jericoacoara site is located on a gently undulating sand sheet approximately 250 meters downwind of a patch of gravel and vegetation and 750 meters from the Atlantic Ocean shoreline under dominantly easterly sea breezes. The Rancho Guadalupe site is located on a flat sand patch approximately 100 meters from a canyon containing the Santa Maria River estuary and 950 meters from the Pacific Ocean shoreline. The Oceano site is located on a gently sloped (up in landward direction) sand patch approximately 300 meters downwind of coastal foredunes and 650 meters from the Pacific Ocean shoreline. Both the Rancho Guadalupe and Oceano sites, separated by about 10 km, are contained within the Guadalupe-Nipomo Dune Complex (Cooper, 1967) shaped by dominantly westerly sea breezes. Additional information about these field sites is provided in Table 1.

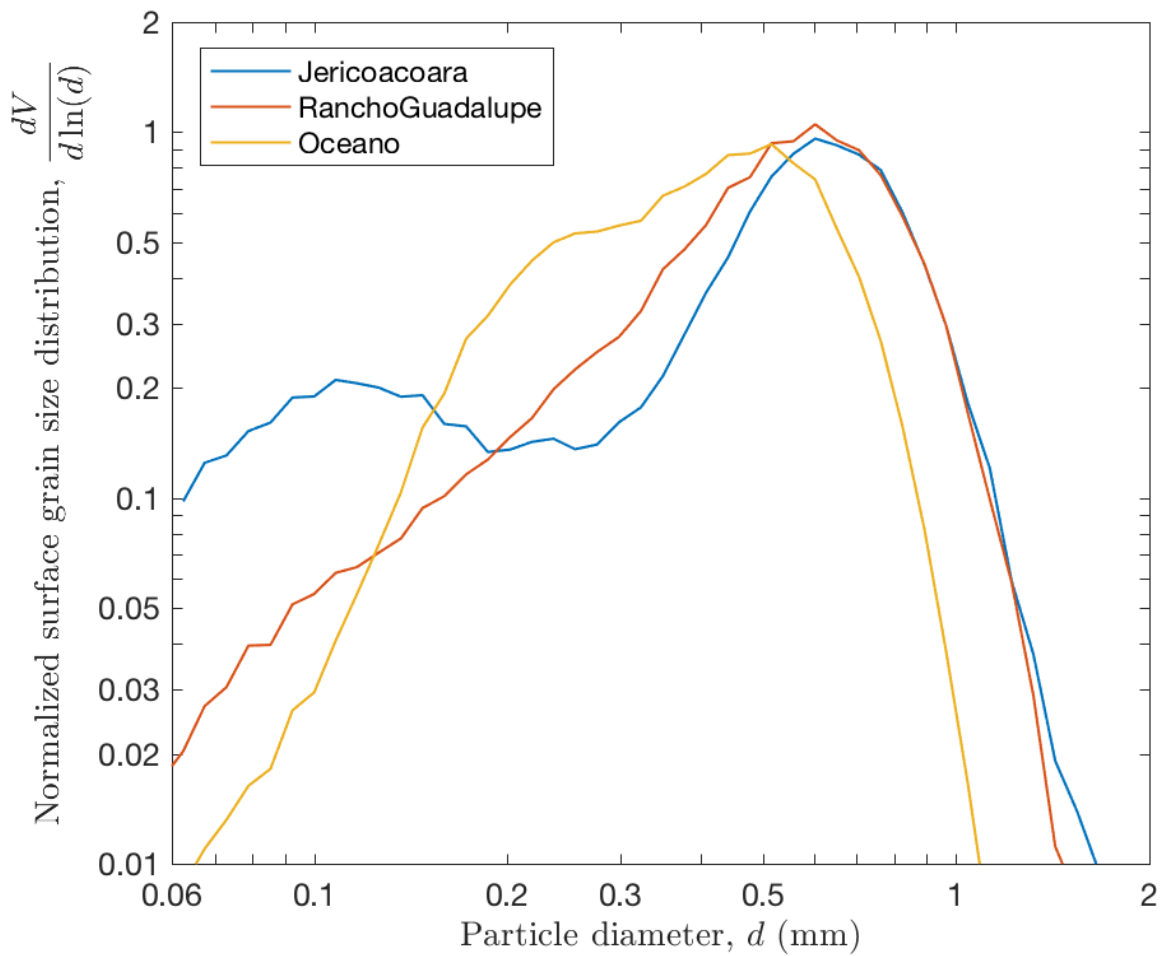

**Figure 1.** Grain size distributions determined from surface samples collected at field sites. $V$ is particle volume and $d$ is particle diameter. Jericoacoara and Rancho Guadalupe share a similar 0.6 mm primary modal grain size but differ in their distribution of fine particle sizes (bimodal at Jericoacoara versus unimodal at Rancho Guadalupe). Oceano surface sand is slightly finer (0.5 mm modal grain size) than at the other two sites.

---

### 2.2. Grain-size distributions

We analyzed grain-size distributions of sand samples collected from the bed surface and from airborne sand traps at each field site. For all of these samples, we obtained particle size distributions by volume using a Retsch Camsizer optical grain size analysis instrument, located at the Sediment Dynamics Laboratory at the University of Pennsylvania (Jerolmack et al., 2011).

To collect the samples, we used a trowel to scrape up the top ~1 cm of surface sand at the beginning and end of each field day. To discern the "typical" grain size distribution from the samples, we averaged all surface grain size distributions into a single representative distribution (Fig. 1). We estimated $10^{th}$, $50^{th}$, and $90^{th}$ percentile index values – $d_{10}$, $d_{50}$, and $d_{90}$, respectively – for each site-averaged grain-size distribution (Table 1). We estimated the corresponding uncertainties in these site-averaged index values as the standard deviation of grain-size index values for individual samples.

We also collected airborne samples in BSNE (Big Springs Number Eight) sand traps (Fryrear, 1986) deployed at various heights above the sand surface. We found no systematic variation of grain size with shear velocity, but we did notice changes in grain size with height above the sand surface. In Sec. 5.4, we consider the effect of vertical variations in the airborne grain size with height to interpret some of the variability in the calibration factors used to convert from relative to absolute saltation fluxes for HF sensors.

## 3. Instrumentation

In this section, we describe the instruments used for characterizing wind and saltation and the associated spatial configurations of these instruments (Table 1). As a convention for describing instrument locations, $x$ is the streamwise coordinate (positive downwind), $y$ is the spanwise coordinate (positive according to right-hand rule, i.e., to the right if looking upwind), and $z$ is the vertical coordinate (positive away from the ground surface). We set the origin ($x, y, z = 0,0,0$) as the sand surface in the center of the HF instrument stand at each site. The $x, y$ coordinate grid remained fixed for each deployment site, while the $z = 0$ level varied through time with sand aggradation and deflation at each site, as we explain further below.

### 3.1. Wind measurements

Sonic anemometers (RM Young 81000 recording at 25 Hz at Jericoacoara and Rancho Guadalupe and Campbell CSAT3 recording at 50 Hz at Oceano) made high-frequency measurements of the three-dimensional wind vector: $u, v, w$, respectively in the streamwise, spanwise, and vertical directions ($x, y, z$). Sonic anemometers also collected simultaneous measurements of equivalent temperature, useful for calculations of atmospheric stability. Though we deployed sonic anemometers at multiple heights (Table 1), we used observations only from the lowest sonic anemometer at each site for the analyses described in this paper. These lowermost anemometers were positioned at a height $z_U \approx 0.5$ m above the ground surface (Table 1). We selected the lowest anemometers as most representative of shear stress at the bed and least likely to be affected by boundary layer instability farther from the surface. In support of this choice to use data only from the lowest anemometers, measurements from anemometers higher above the surface yield wind shear stresses similar to those measured by the lowest anemometer (Martin and Kok, 2017a).

### 3.2. Saltation flux measurements

We used multiple instruments to characterize sand flux. BSNE sand traps at multiple heights above the surface provided absolute LF measurements of saltation flux, but with low time resolution. Wenglor optical particle counters provided HF relative time series of saltation counts, but required calibration to determine absolute fluxes. Downward-facing laser distance sensors measured local bed elevation to determine changes in saltation sensor heights through time, thus aiding the HF saltation flux calibration procedure.

In our subsequent descriptions, we use the subscript "LF" to specify low-frequency measurements and calculations from the BSNEs, and the subscript "HF" to specify high-frequency measurements from the Wenglors. We adopt these subscripts with the understanding that such measurements and calculations could also be made for other types of LF traps and HF sensors.

#### 3.2.1. BSNE sand traps for low-frequency (LF) measurements

BSNE sand traps (Fryrear, 1986) collected airborne saltating sand over fixed time intervals at multiple heights. As we further discuss in Sec. 5.2.1, BSNEs are well calibrated and provide high collection efficiency (Goossens et al., 2000). Most of the BSNEs were the standard type, with vertical opening size $H_{LF} = 5$ cm and horizontal opening size $W_{LF} = 2$ cm. The one exception was one 'modified' BSNE with $H_{LF} = 1$ cm, which we deployed at the Oceano site to measure saltation flux close to the surface.

BSNEs were attached to wind vanes on vertical poles, allowing them to rotate freely with the prevailing wind direction. We placed all BSNEs at the same streamwise position, $x = 0$, corresponding with all other instruments. To increase the vertical ($z$) resolution of the BSNE profile, we placed BSNEs at multiple spanwise ($y$) locations, because trap bulkiness made close vertical spacing of all BSNEs at the same $y$ position impossible. At the beginning and end of each field day, we measured the vertical distance from the ground to the bottom of the trap opening $z_{bot,LF,i}$ for each trap $i$.

We performed BSNE trap collections for specified time intervals $T_{LF}$, usually one hour. Occasionally, we modified $T_{LF}$ to ensure that traps collected a reasonable amount of sand: we extended $T_{LF}$ during periods of very weak saltation and shortened $T_{LF}$ during periods of intense saltation. We weighed all BSNE samples in the lab using a SmartWeigh Pro Pocket Scale TOP2KG (accuracy $\pm 0.1$g) to determine saltation masses $m_{LF,i}$ for each BSNE and each time interval.

#### 3.2.2. Wenglor optical particle counters for high-frequency (HF) measurements

Wenglor optical particle counters (model YH03PCT8 with 3 cm path length) detected passage of individual saltators at multiple heights $z_{HF,i}$ above the bed surface. Wenglor sensors then transmitted these detections as pulses to a Campbell CR1000 datalogger recording at 25 Hz, thus providing HF measurements of saltation pulse counts $n_{HF,i}$ for each sensor $i$. Due to frequent blockage of Wenglor lenses by dust, which manifested as abrupt unexplained changes in the $n_{HF,i}$ time series, we often excluded specific sensors from our analyses. The number of HF sensors included in analyses therefore varied from 3 to 9. Time intervals with fewer than 3 Wenglors were excluded as insufficient for fitting saltation flux profiles.

### 3.2.3. Laser distance sensors for bed elevation

Laser distance sensors (Sick DT35), co-located with the Wenglor laser particle counters, were pointed downward toward the surface to measure changes of bed elevation through time. Based on these distances $z_{dist}$ and known relative vertical distances between each Wenglor $i$ and the distance sensors $z_{rel,i}$, we estimated absolute elevations for each Wenglor $z_{HF,i}$. We describe specifics of this Wenglor height calculation in Sec. 4.2.3.

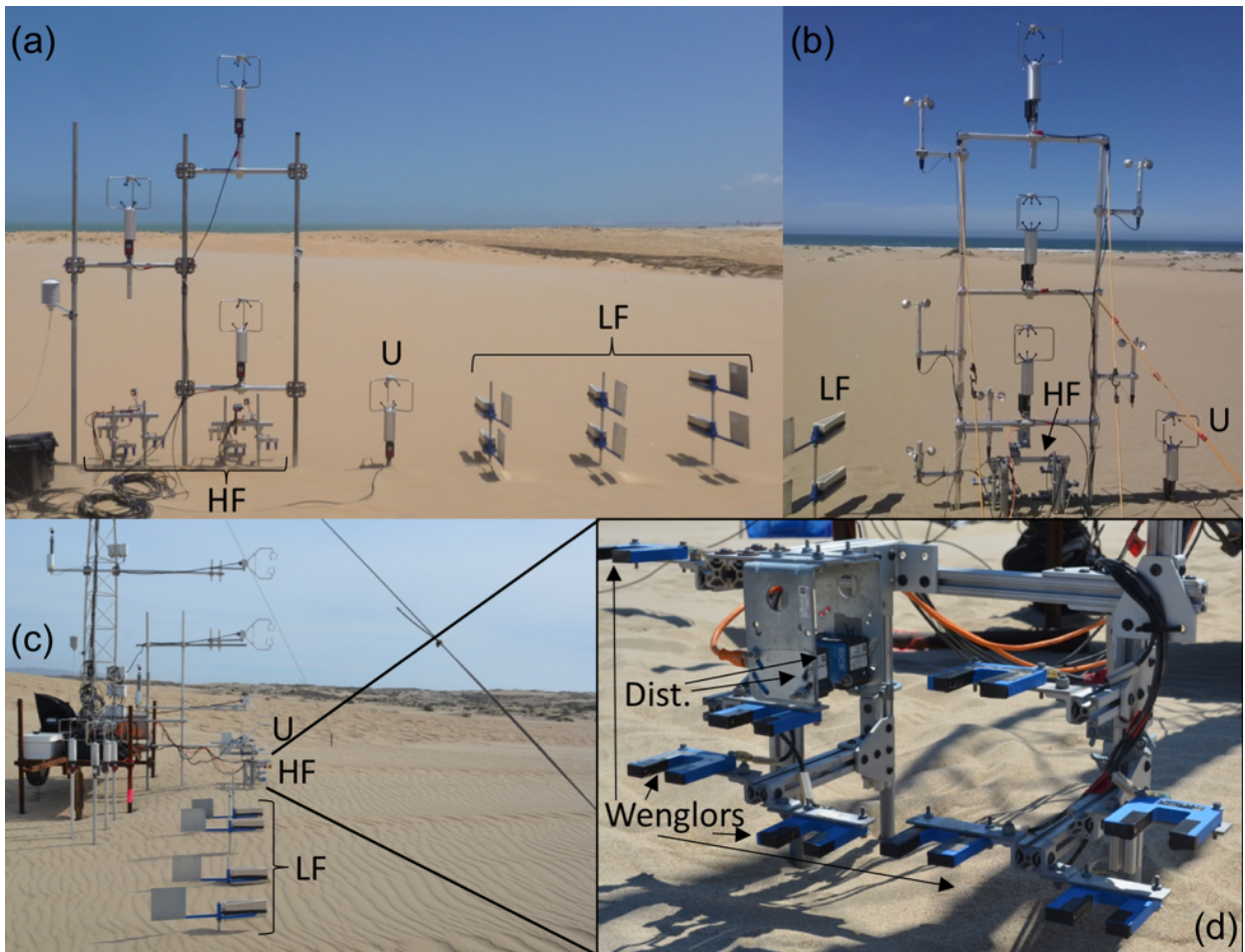

**Figure 2.** Photos of field sites, with locations of LF saltation traps, HF saltation sensors, and sonic anemometers indicated for reference. At each site, only the lowermost sonic anemometer

(denoted "U") was included in the analysis. (a) Jericoacoara, looking upwind. LF traps were separated into 3 towers, and HF sensors were divided into 2 arrays. (b) Rancho Guadalupe, looking upwind. Some LF traps are not shown in the figure. All HF sensors were included in a single array. (c) Oceano, looking from the side, with $+y$ LF traps in foreground and HF sensors in the background. (d) Close-up of Wenglor and distance sensor array for HF measurements at Oceano, looking downwind.

### 3.3. Instrument layouts

Instrument layouts varied from site to site, as illustrated in Fig. 2. We detail the layouts of specific instrument types below.

At all field sites, measurements from the sonic anemometer, BSNE traps, and Wenglor sensors were co-located in the streamwise direction at $x = 0$. In the spanwise ($y$) direction, we positioned the BSNE traps away from the Wenglors and sonic anemometer to reduce interference among instruments. At Jericoacoara, we placed all BSNEs on one side ($+y$) of the Wenglors and anemometer (Fig. 2a). At Rancho Guadalupe and Oceano, we placed BSNEs on both sides of the setup ($\pm y$) (Fig. 2b), though BSNEs were more widely spaced at Oceano ($y \approx \pm 5$ m) (Fig. 2c). At Jericoacoara and Rancho Guadalupe, Wenglors were also separated in the $y$-direction from sonic anemometers (Fig. 2a and b), whereas the sonic anemometer and Wenglors shared the same $y \approx 0$ at Oceano (Fig. 2c).

In addition to the spanwise separation of different measurements types, we also separated individual BSNE traps and Wenglor sensors in the $y$-direction to accommodate the spatial footprints of individual instruments. The maximum spanwise BSNE separations were 1.5 m, 3 m, and 10 m at Jericoacoara, Rancho Guadalupe, and Oceano, respectively. We adopted a much wider BSNE spanwise separation at Oceano to reduce potential instrument interference issues at this site. The maximum spanwise Wenglor separations were 1.2 m, 0.6 m, and 0.6 m at Jericoacoara, Rancho Guadalupe, and Oceano, respectively.

For practical reasons, all Wenglors and laser distance sensors were mounted onto fixed stands, unable to rotate with the wind like the BSNE traps. As wind direction during saltation varied within a narrow range of $\pm 15°$ (Martin and Kok, 2017a), we expect that the combination of rotating LF traps and rigid HF sensors had only a small effect on our analyses, and these variations were likely capture in the calibration factors (Sec. 4.3 below). At Jericoacoara, we mounted Wenglors in a vertical orientation with prongs facing downward, and these Wenglors were mounted on two separate stands (Fig. 2a). At Rancho Guadalupe and Oceano, we mounted Wenglors horizontally with prongs facing upwind, and these Wenglors were mounted on a single stand (Fig. 2b-d). We mounted distance sensors at known relative heights $z_{rel,i}$ directly on each Wenglor stand, providing height calibration for all Wenglors on each stand. Fig. 2d provides a close-up view of the Wenglor and distance sensor mounting at Oceano.

### 4. Methodology and illustrative results

In this section, we explain the four-step methodology for obtaining calibrated, high-resolution measurements of the total (vertically-integrated) saltation flux. The sequence of steps is as follows: (1) perform exponential fits to LF trap vertical profiles of saltation flux, (2) determine the empirical calibration factors through comparison of LF exponential fits to HF number counts over concurrent time intervals, (3) apply these calibration factors to subsamples of the saltation count time series to obtain HF height-specific saltation fluxes, and (4) aggregate the calibrated HF height-specific saltation fluxes into total saltation fluxes through application of profile fitting and summation techniques. An overview of these methods is shown in Fig. 3. In addition, we provide a glossary of all variables used to describe this methodology in Supporting Information Sec. S2.

**4.1. Evaluating low-frequency (LF) saltation flux profiles from BSNE traps**
In this section, we describe methods for obtaining LF vertical profiles of horizontal saltation flux measured with BSNE sand traps. We compare the quality of power law versus exponential fits to LF saltation flux profiles. In showing the superiority of exponential profile fits, we justify the application of exponential fitting in subsequent sections of this paper. Though this section specifically describes measurements obtained from BSNE sand traps, we describe our methods in a way that is meant to be generally applicable to all vertical profiles of LF saltation measurements, regardless of trap type.

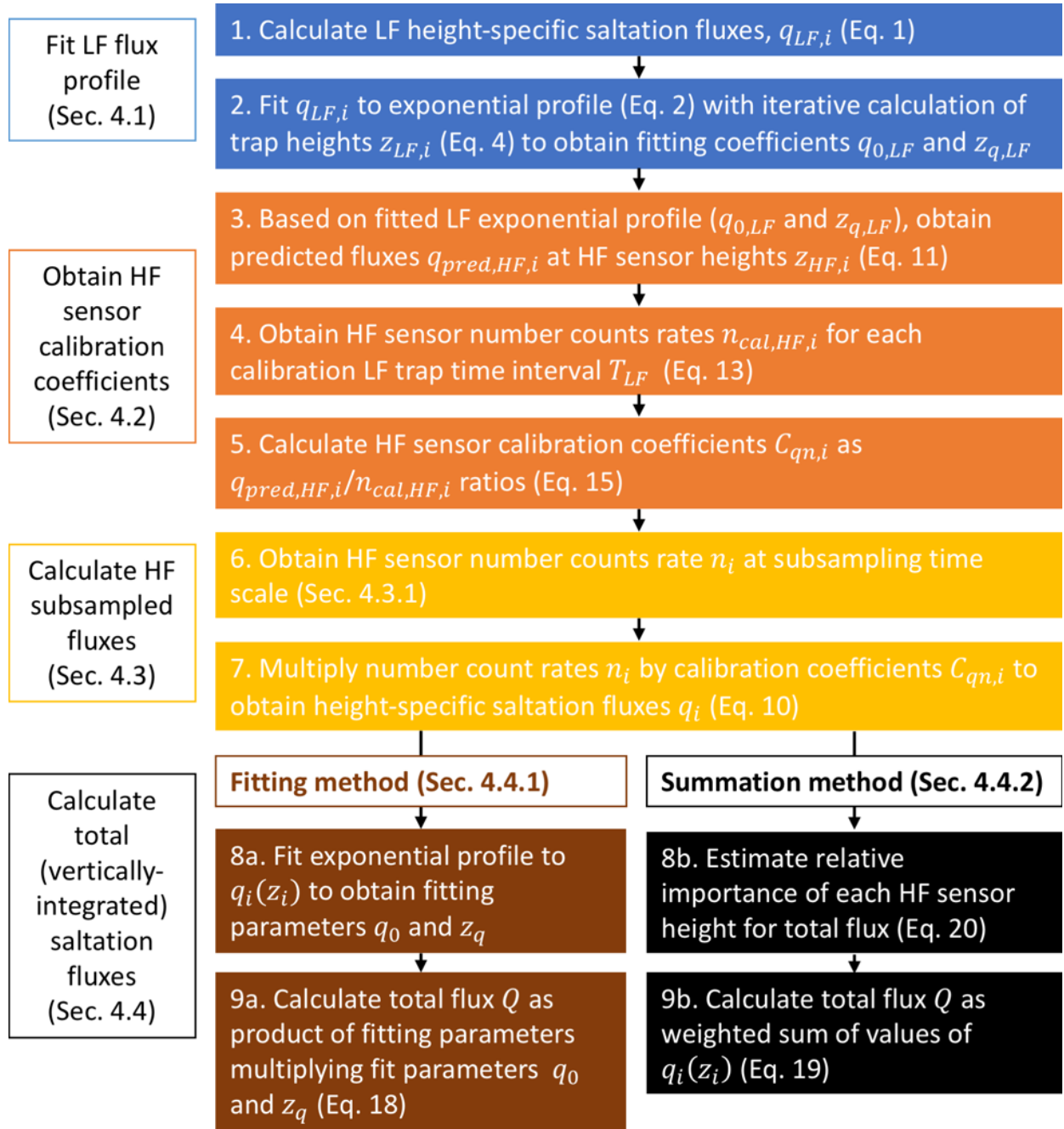

**Figure 3.** Flow chart overview of methodology for obtaining the total (vertically-integrated) saltation flux from the LF and HF measurements.

### 4.1.1. Calculating LF height-specific saltation fluxes

We calculate the LF height-specific horizontal saltation flux $q_{LF,i}$ for the $i$th trap based on the measurement interval of duration $T_{LF}$ (typically 1 hour, see Sec. 3.2.1), the trap opening height $H_{LF}$, trap width $W_{LF}$, and total mass of sand collected $m_{LF,i}$:

$$q_{LF,i} = \frac{m_{LF,i}}{T_{LF}H_{LF}W_{LF}}. \quad (1)$$

The parameters describing trap dimensions are illustrated in Fig. 4. We compute flux uncertainty $\sigma_{q_{LF,i}}$ through propagation of the quantities included in the calculation (see Supporting Information S1.4).

**4.1.2. Exponential fits to LF flux profiles**
In this subsection, we describe methods for performing exponential fits to LF saltation flux profiles. We perform exponential fits to vertical profiles of $q_{LF,i}$, which take the form (Supporting Information S1.3):

$$q_{exp,LF,i} = q_{0,LF} \exp\left(-\frac{z_{LF,i}}{z_{q,LF}}\right), \quad (2)$$

where $q_{exp,LF,i}$ is the exponential best-fit saltation flux for trap $i$, $z_{LF,i}$ is the height of trap $i$, $q_{0,LF}$ is the best-fit scaling parameter, and $z_{q,LF}$ is the best-fit $e$-folding saltation layer height for the profile.

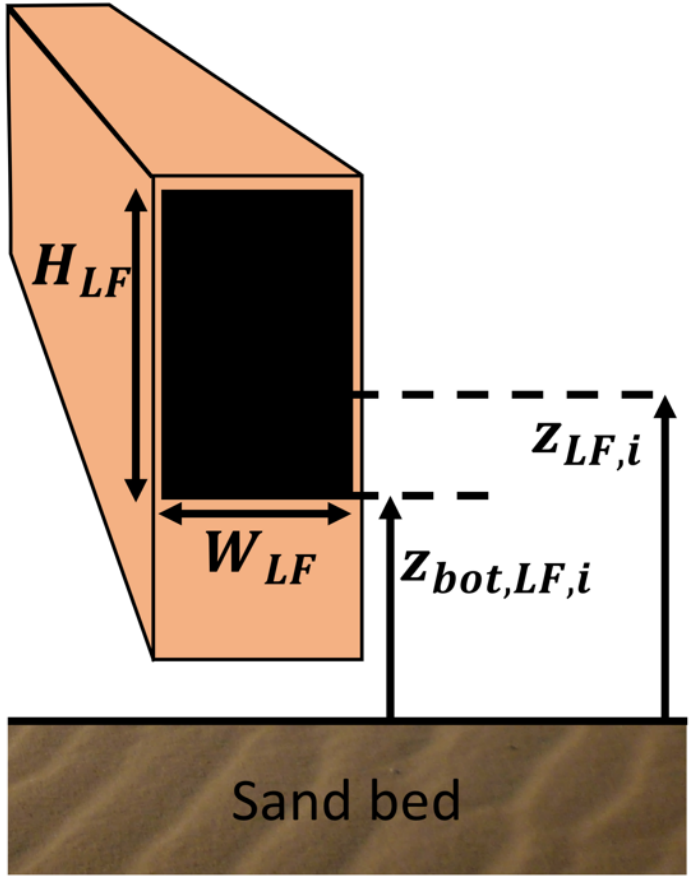

**Figure 4.** Illustration of variables describing the geometry of the LF trap opening and positioning for calculations in Sec. 4.1. $H_{LF}$ is trap opening height, $W_{LF}$ is trap width, $z_{bot,LF,i}$ is the bottom elevation of the trap opening, and $z_{LF,i}$ is the calculated representative height of the trap opening.

---

Prior to fitting, values for $z_{LF,i}$ are not available; we have only the heights of the bottoms of the traps $z_{bot,LF,i}$ and the associated trap opening sizes $H_{LF}$. The "true" value for $z_{LF,i}$ is between $z_{bot,LF,i}$ and $z_{bot,LF,i} + H_{LF}$ (Fig. 4), and it represents the height at which the best-fit local flux $q_{exp,LF,i}$ equals the fit-predicted height-integrated flux for the trap, i.e.:

$$q_{exp,LF,i}(z_{LF,i}) = \frac{1}{H_{LF}} \int_{z_{bot,LF,i}}^{z_{bot,LF,i}+H_{LF}} q_{exp,LF,i}(z)\, dz. \quad (3)$$

Plugging Eq. 2 into Eq. 3 and simplifying gives:

$$z_{LF,i} = -z_{q,LF} \log\left(-\frac{z_{q,LF}}{H_{LF}} \exp\left(-\frac{z_{bot,LF,i}}{z_{q,LF}}\right)\left(\exp\left(-\frac{H_{LF}}{z_{q,LF}}\right) - 1\right)\right). \quad (4)$$

Examination of Eq. 4 reveals that calculation of $z_{LF,i}$ depends on knowledge of $z_{q,LF}$. However, $z_{q,LF}$ conversely requires values of $z_{LF,i}$ for fitting to Eq. 2. To address this issue, we iteratively fit the exponential profile (Eq. 2) and calculate LF trap heights (Eq. 4) until the best-fit values $q_{0,LF}$ and $z_{q,LF}$ converge. For each exponential fit, we apply the fitting methods described in Supporting Information S1.3 to the values of $q_{LF,i}$ and $\sigma_{q_{LF,i}}$ from Sec. 4.1, the iterative $z_{LF,i}$, and the assumed height uncertainties $\sigma_{z_{LF,i}} = 0.5$ cm for all LF traps. Each exponential fit iteration yields best-fit values $q_{0,LF}$ and $z_{q,LF}$, with respective uncertainties, $\sigma_{q_{0,LF}}$ and $\sigma_{z_{q,LF}}$. A typical exponential fit to an LF flux profile is shown in Fig. 5.

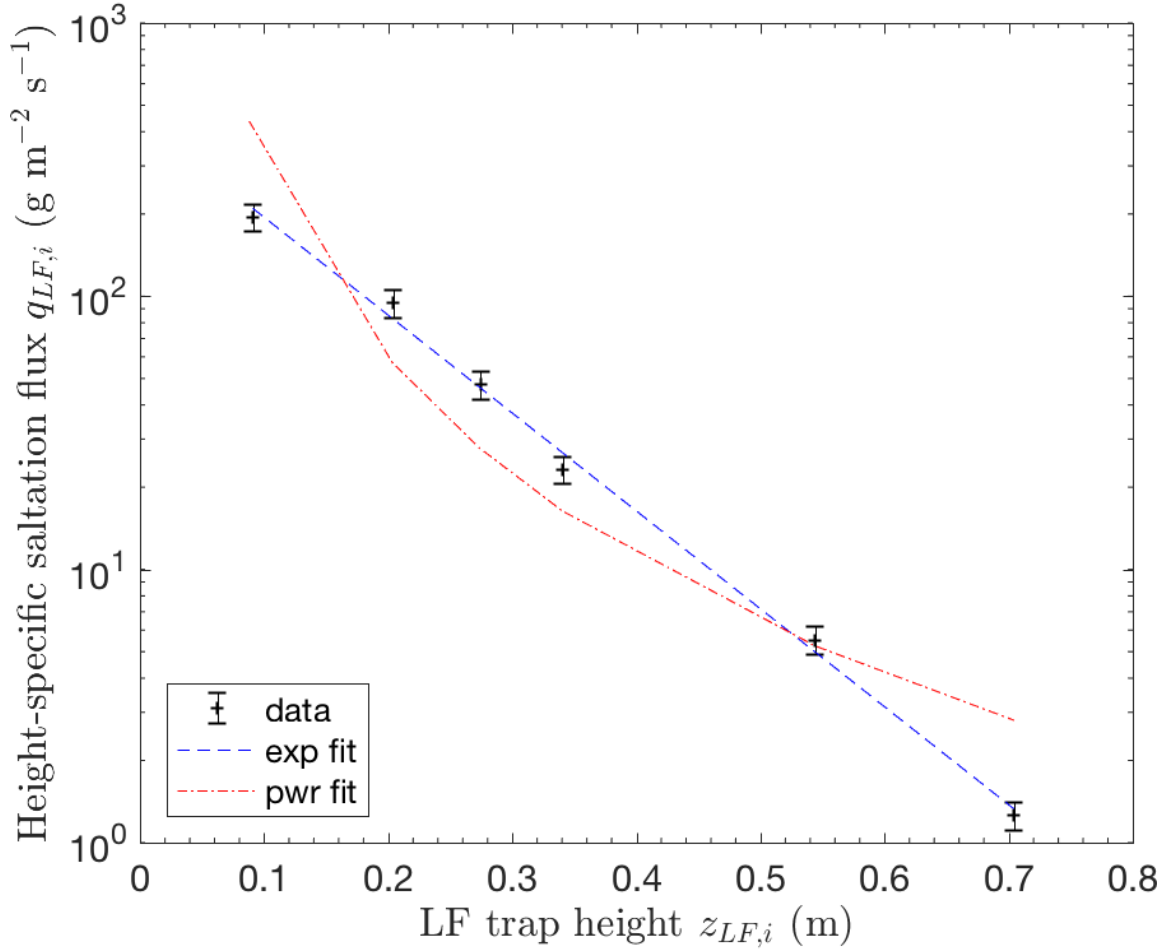

**Figure 5.** Representative LF flux profile from Rancho Guadalupe (24 March 2015, 15:49-16:34 local time). Data points show LF height-specific horizontal saltation flux values $q_{LF,i}$ (Eq. 1) and their uncertainties (Supporting Information Eq. S24) versus trap heights $z_{LF,i}$. Blue dashed line is exponential fit to profile (Eq. 2). Red dot-dashed line is the power law fit to this profile (Eq. 5). Reduced chi-squared values are $\chi_\nu^2 = 1.1$ and $\chi_\nu^2 = 64.9$ for the exponential and power law fits, respectively.

---

We note that our method of trap height estimation differs from the protocol described by Ellis et al. (2009a) for estimating trap height from the geometric mean of $z_{bot,LF,i}$ and $z_{bot,LF,i} + H_{LF}$. In most cases, our calculated values of $z_{LF,i}$ are statistically indistinguishable from those that would have been obtained by the geometric mean method. However, for traps very close to the bed, we do find that the Ellis et al. method slightly underestimates the $z_{LF,i}$ value (See Supporting Information Fig. S1), indicating the superiority of the iterative method for general use.

### 4.1.3. Power-law fits to LF flux profiles

In this subsection, we describe methods for performing power law fits to LF saltation flux profiles. As described further in Supporting Information S1.4, power law fits take the form

$$q_{pwr,LF,i} = q_{p,LF} z_{LF,i}^{-k_{z,LF}}, \qquad (5)$$

where $z_{LF,i}$ are the heights of individual LF traps, and $q_{p,LF}$ and $k_{z,LF}$ are respectively the best-fit scaling parameter and characteristic power law exponent for the LF profile.

As with the exponential fitting, values for $z_{LF,i}$ are not known *a priori*. As before, we therefore also estimate the "true" value for $z_{LF,i}$ as the height at which the best-fit local flux equals the fit-predicted height-integrated flux for the trap. Adapting Eq. 3 to the power law profile (Eq. 5) gives:

$$z_{LF,i} = \left( \frac{1}{(1-k_{z,LF})H_{LF}} \left[ \left(z_{bot,LF,i} + H_{LF}\right)^{1-k_{z,LF}} - \left(z_{bot,LF,i}\right)^{1-k_{z,LF}} \right] \right)^{-1/k_{z,LF}}. \qquad (6)$$

As for the exponential profiles, determination of the $z_{LF,i}$ in Eq. 6 requires iterative calculations of $k_{z,B}$ by fitting to Eq. 5. For each power law fit, we apply the fitting methods described in Supporting Information S1.4 to the values of $q_{LF,i}$ and $\sigma_{q_{LF,i}}$ the iterative $z_{LF,i}$, and the assumed height uncertainties $\sigma_{z_{LF,i}} = 0.5$ cm for all LF traps. Each power law fit iteration yields best fit values $q_{p,LF}$ and $k_{z,LF}$, with respective uncertainties, $\sigma_{q_{p,LF}}$ and $\sigma_{k_{z,LF}}$. Fig. 5 displays a typical power law fit to an LF flux profile.

### 4.1.4. Comparison of exponential and power-law fits
To judge the favorability of exponential versus power law fits to the LF flux profiles, we compute the $\chi^2$ statistic for each profile fit, which we calculate as:

$$\chi_v^2 = \frac{1}{v} \sum_i \frac{(q_{LF,i} - q_{fit,LF,i})^2}{\sigma_{q_{LF,i}}^2}, \qquad (7)$$

where $q_{fit,LF,i}$ are the fit-estimated fluxes for the exponential (Eq. 2) and power law (Eq. 5) profile fits, and $v$ is the number of data points minus the number of fitting parameters (2 for both the exponential and power law fits). For the example saltation flux profile shown in Fig. 5, the $\chi_v^2$ value is substantially smaller for the exponential fit than for the power law fit, indicating the superiority of the exponential. Beyond this specific example, $\chi_v^2$ values for exponential fits are generally lower than for power law fits – at all three field sites and over a range of saltation intensities (Fig. S2). Therefore, we use the exponential model from this point forward to characterize saltation flux profiles.

### 4.1.5. Saltation profile fits for widely spaced traps
For measurements from Oceano, we modify the LF fitting procedure to account for the spanwise separation of LF traps, which was substantial at this site (about 9 m between $+y$ and $-y$ LF traps). As a result of this separation, which covered a measurable gradient in surface grain size properties, the $+y$ and $-y$ LF profiles show significant difference (Supporting Information Fig. S3).

We therefore choose to fit separate profiles to the $+y$ and $-y$ LF trap profiles, then to combine $+y$ and $-y$ LF exponential profile fit pairs into single values for each LF measurement interval at Oceano. We do this because HF sensor arrays were located roughly halfway between $+y$ and $-y$ LF profiles. Denoting the separate LF profile fit values as $z_{q,LF,+}$ and $q_{0,LF,+}$ for the $+y$ LF profile fits and $z_{q,LF,-}$ and $q_{0,LF,-}$ for the $-y$ LF profile fits, we calculate combined LF profile fit values at Oceano as:

$$z_{q,LF} = \frac{z_{q,LF,+} + z_{q,LF,-}}{2}, \tag{8}$$

i.e., the arithmetic average, because the $z_{i,LF}$ are computed in linear space, and

$$q_{0,LF} = \sqrt{q_{0,LF,+} q_{0,LF,-}}, \tag{9}$$

i.e., the geometric average, because the $q_{i,LF}$ are computed in logarithmic space. We describe methods for computing the combined LF profile fit value uncertainties, $\sigma_{z_{q,LF}}$ and $q_{0,LF}$, in Supporting Information S1.5.

### 4.2. Calibrating high-frequency (HF) saltation mass fluxes

In this section, we describe methods for using the absolute mass fluxes obtained from LF flux profiles to calibrate HF sensor counts, here collected with Wenglor optical particle counters sampling at 25 Hz. We first present an overview of the calibration method, then we detail quality control criteria for selection of HF sensors, and finally we provide details on each individual calibration step.

### 4.2.1. Overview of calibration method

An overview of the three primary steps in the calibration method is provided in Fig. 6. First, we extrapolate LF profile exponential fits for each measurement time interval to predict height-specific saltation fluxes $q_{pred,HF,i}$ for each HF sensor $i$ at height $z_{HF,i}$ over this same time interval (Fig. 6a). Second, we determine time-averaged count rates $n_{cal,HF,i}$ for each HF sensor over these calibration time intervals (Fig. 6b). Third, we compute the ratio of $q_{pred,HF,i}$ and $n_{cal,HF,i}$ to obtain calibration factors $C_{qn,i}$ to convert from number counts to saltation fluxes for each HF sensor over time intervals of arbitrary duration $\Delta t$ (Fig. 6c), as:

$$q_i = C_{qn,i} n_{HF,i}, \tag{10}$$

where $q_i$ is the calibrated saltation flux and $n_{HF,i}$ is the HF sensor count rate over $\Delta t$. This calibration method is simpler and more practical than the method proposed by Barchyn et al. (2014a), as it does not depend on grain-size measurements at HF sensor heights (see further discussion in Sec. 4.2.6). However, it does rely upon an assumption of linear response between sensor detected particle counts and saltation flux. The experiments of Barchyn et al. (2014a) support this linear response assumption for the Wenglor sensors used here. We further discuss the selection of HF sensors in Sec. 5.2.2.

We note that HF sensor calibration factors varied substantially through time, due to variations in sensor performance with scratches and buildup of dust and condensation on lenses. Subtle changes in wind direction, which affected the alignment of fixed HF sensors with rotating LF sensors, may have also contributed to variations in calibration factors. Our calibration method is thus able to reasonably account for such variation in sensor performance and alignment. However, in instances of unusual HF sensor performance, we decide to exclude certain sensors from calibration and subsequent analysis. Before detailing the three steps in the calibration method, we first describe our quality control procedures for determining which HF sensors to consider for analysis.

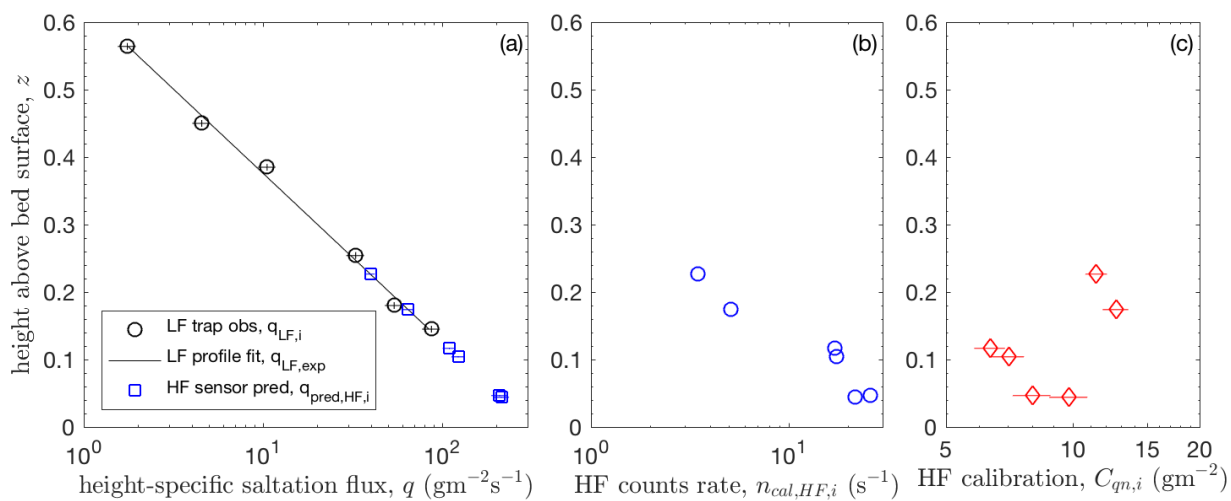

**Figure 6.** Demonstration of the three primary steps in the method for determining HF sensor calibration factors, as laid out in detail in Sec. 4.2 of this paper. The data are from Jericoacoara on 14 November 2014, from 15:26 to 16:26 local time. (a) First, we predict the saltation fluxes at the heights of the HF sensors. To do this, we obtain the exponential fit $q_{exp,LF}$ (Eq. 2 – black line) to LF trap saltation fluxes $q_{LF,i}$ versus trap heights $z_{LF,i}$ (black circles), then we extrapolate from this exponential fit to get expected saltation fluxes at HF sensor heights $q_{pred,HF,i}$ (Eq. 11 – blue squares). (b) Second, we obtain the corresponding HF sensor number count rate over the calibration interval $n_{cal,HF,i}$ (Eq. 13 – blue circles). (c) Third, we compute the ratio of $q_{pred,HF,i}$ (panel a) and $n_{cal,HF,i}$ (panel b) to obtain the calibration factors for HF sensors $C_{qn,i}$ (Eq. 14 – red diamonds).

### 4.2.2. HF sensor quality control

To determine which HF sensors to include in the calibration process for each measurement time interval, we consider two criteria: (1) the height of the sensor above the surface, and (2) the stability of the saltation number counts time series in comparison to other sensors. For the first criterion, we set a lower limit HF sensor height $z_{HF,i} = 1.8$ cm, below which we exclude sensors from analysis. This height represents the minimum distance of the sensor from the ground, such that the bottom of the instrument (in our case, the Wenglor optical particle counter) is separated from the ground by at least one instrument width. Because of the effects of changing bed elevation due to migrating ripples and scour, we found that relative uncertainties of HF sensor heights for $z_{HF,i} < 1.8$ cm were too large for inclusion in the analysis.

To evaluate the second criteria, we perform a qualitative analysis comparing time series of detected saltation particle counts across all sensors to diagnose drift in sensor performance. Due to differences in sensor heights and sensitivities, we expect substantial differences in long-term mean particle count rates among sensors. However, we also expect that these differences among sensors remain roughly stable through time – that each sensor displays roughly corresponding fluctuations in particle counts through time. In cases where one or more sensors displays sudden changes or significant long-term trends not reflected in saltation count time series from other sensors, we consider it likely that the sensor performance was degraded by dust build-up on the lens or other factors causing a slow decline in count rate, or a sudden inability to count particles when the laser is blocked (Hugenholtz and Barchyn, 2011; Barchyn et al., 2014a). Therefore, we exclude such malfunctioning sensors from subsequent calibration and analysis. This quality control protocol is an evolved version of the recommendations of Barchyn et al. (2014a), who suggest that all Wenglor sensors should be deployed in pairs to provide cross-checks on sensor counts.

The HF sensor calibration methods and subsequent flux calculations described below apply only to those sensors that fulfilled our two quality control criteria: high enough off the ground (i.e., $z_{HF,i} > 1.8$ cm) and relatively unaffected by long-term drift or sensor failure.

### 4.2.3. Determination of predicted saltation fluxes at HF sensor heights

For each functioning HF sensor *i* and each LF trap measurement interval, we calculate an expected height-specific horizontal saltation flux $q_{pred,HF,i}$ at sensor height $z_{HF,i}$ based on the LF exponential profile fit parameters $q_{0,LF}$ and $z_{q,LF}$ for that time interval (Fig. 6a):

$$q_{pred,HF,i} = q_{0,LF} \exp\left(-\frac{z_{HF,i}}{z_{q,LF}}\right). \tag{11}$$

We describe methods for computing the associated uncertainty in expected saltation flux $\sigma_{q_{pred,HF,i}}$ in Supporting Information S1.6.

We calculate the height of each HF sensor $z_{HF,i}$ as the sum of the height of the HF sensor relative to the distance sensor $z_{rel,HF,i}$ and the height measured by the distance sensor of its elevation with respect to the bed surface $z_{dist}$:

$$z_{HF,i} = z_{rel,HF,i} + z_{dist}. \tag{12}$$

This method of computing HF sensor height is illustrated in Fig. 7. Because HF sensors and distance sensors were mounted on a fixed stand, relative height $z_{rel,HF,i}$ remains constant through each deployment day. In contrast, $z_{dist}$ captures variability in bed surface elevation, associated either with bed surface change (i.e., migration of ripples or bed scour) or gradual settling of the mounting stand, which may have occurred after the initial installation of instruments at each field site. However, because saltation trajectories originated from a variety of locations upwind of the HF sensors, we do not consider instantaneous variations of $z_{dist}$ due to the migration of individual ripples; rather, we compute $z_{dist}$ as a single time-averaged value for each discrete period of HF sensor deployment. It is likely, though, that shorter-term variations in bed elevation could have affected fluctuations in creep and reptation flux (Bauer and Davidson-Arnott, 2014; Durán et al., 2014); thus, the $z_{dist}$ time series could be useful for quantifying these non-saltation contributions to the total aeolian sand transport in future work.

### 4.2.4. Determination of saltation number counts for HF sensors

As the next step in the HF sensor calibration, we determine the particle count rate $n_{cal,HF,i}$ for each HF sensor $i$ over the corresponding LF trap measurement interval of duration $T_{LF}$:

$$n_{cal,HF,i} = N_{cal,HF,i}/T_{LF}, \qquad (13)$$

where $N_{cal,HF,i}$ is the total number of particles counted by HF sensor $i$ during $T_{LF}$.

### 4.2.5. Calculation of HF sensor saltation flux calibration factor

Based on the expected height-specific horizontal saltation flux $q_{pred,HF,i}$ and number count rate $n_{cal,HF,i}$ for each HF sensor $i$, we estimate the calibration factor for each HF sensor as:

$$C_{qn,i} = \frac{q_{pred,HF,i}}{n_{cal,HF,i}}. \qquad (14)$$

Methods for estimating the associated calibration factor uncertainty $\sigma_{C_{qn,i}}$ are described in Supporting Information S1.7.

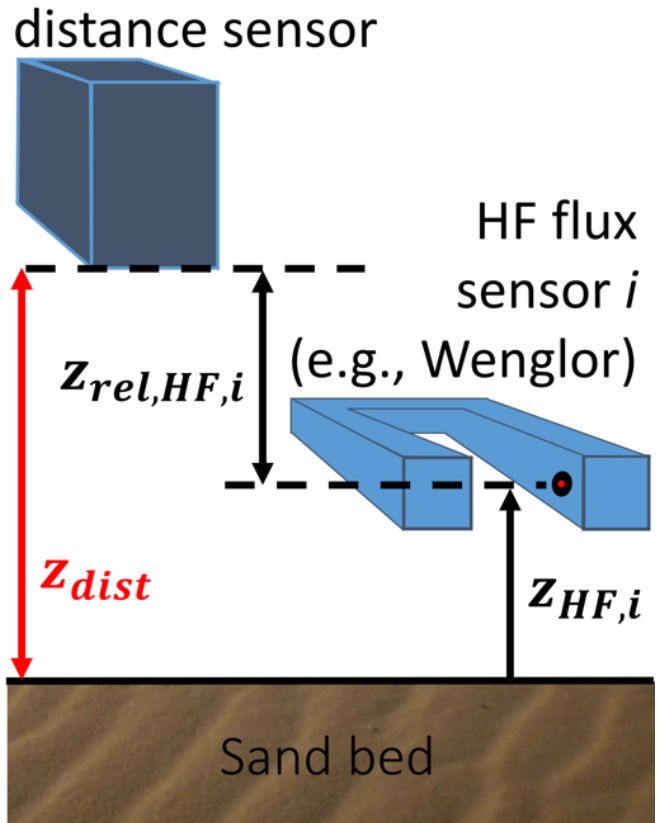

**Figure 7.** Illustration of variables describing the geometry of HF flux sensor positioning relative to the distance sensor and sand bed for calculations in Sec. 4.2. $z_{dist}$ is the measured height of the distance sensor relative to the bed, $z_{rel,HF,i}$ is the height of the HF flux sensor laser relative to the distance sensor, and $z_{HF,i}$ is the resulting calculated HF sensor height (Eq. 12). In this figure, a Wenglor optical particle counter is shown to illustrate the HF sensor geometry.

### 4.2.6. Evaluation of flux calibration factors

To evaluate our calibration method, we first perform a qualitative examination of variation in calibration factors through time. We consider measurements from a single day (3 June 2015) at a single site (Oceano) for which we have a large number of LF trap profiles (8), a wide range of saltation conditions, and a long continuous record of HF sensor measurements (>7h). For these sample measurements, Fig. 8 shows that there is substantial variation in calibration factors across sensors and time intervals throughout the day. Much of the variation in $C_{qn,i}$ appears to be correlated among all sensors, but such synchronous variations appear unrelated to variations in saltation flux measured by the LF traps. It is possible that, by neglecting high-frequency variations in the bed elevation, we are failing to account for systematic variations in the HF sensor heights, which could drive synchronous variations in $C_{qn,i}$ for all sensors.

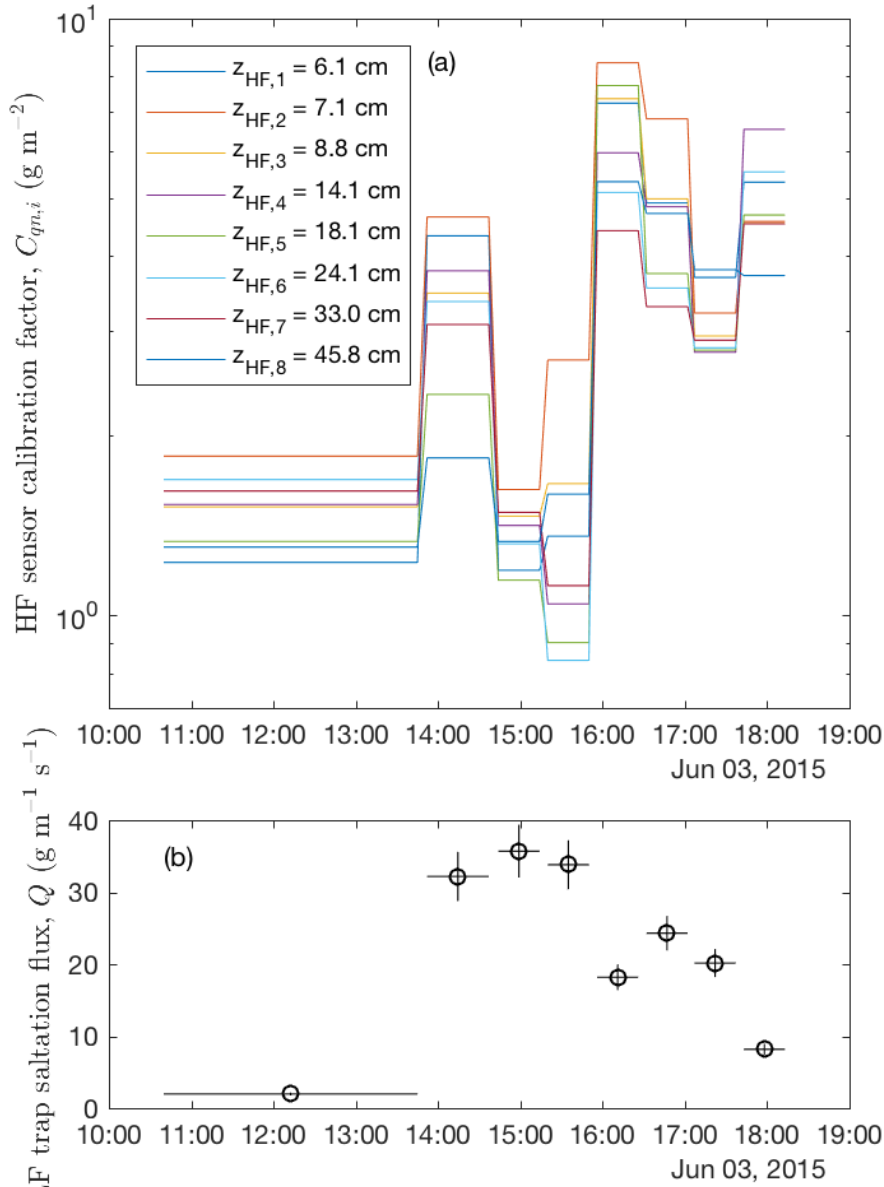

**Figure 8.** (a) Example of variation through time in calibration coefficients $C_{qn,i}$ for HF sensors $i$ at heights $z_{HF,i}$ above the bed surface, calculated for 3 October 2017 at the Oceano field site. Values of $C_{qn,i}$ adjust over discrete time intervals associated with LF trap saltation measurements. (b) Variation through time in the total saltation flux $Q$ for LF saltation trap profiles. Horizontal bars indicate LF trap measurement time intervals. $Q$ values for each of these intervals are calculated through application of the fitting method (described in Sec. 4.4.1 below) to vertical profiles of LF trap height-specific fluxes $q_{LF,i}$; vertical bars indicate uncertainties in $Q$ values for these intervals.

Additional, but less substantial, variation in $C_{qn,i}$ appears related to variation among individual sensor heights. We expect that some of the variation in $C_{qn,i}$ could be explained by variation of airborne particle sizes, because pulse counts from coarser particles represent a greater saltation mass flux than those from finer particles. In particular, assuming spherical particles, we expect that the calibration factor $C_{qn,pred,i}$ for trap $i$ will be:

$$C_{qn,pred,i} = \frac{\frac{\pi}{6}\rho_s \bar{d}^3}{A_{HF}}. \tag{15}$$

where $\rho_s$ is particle density, assumed as 2650 kg/m$^3$ for natural sand, $\bar{d}$ is the volume-weighted mean diameter of particles passing through the HF sensor, and $A_{HF}$ = 18 mm$^2$ is the area of the Wenglor opening, approximated as the product of the sensor's 30 mm path length and 0.6 mm laser beam diameter. In general, we observe a subtle increase in $\bar{d}$ with height above the bed surface; thus, we also expect $C_{qn,pred,i}$ to increase with height.

Both the observed ($C_{qn,i}$; Fig. 9a) and predicted ($C_{qn,pred,i}$; Fig. 9b) calibration factors show a weak increase with height, lending support to the grain-size dependence of the calibration factors, similar to that discussed by Barchyn et al. (2014a). Furthermore, when directly comparing observed and predicted calibration factors (Fig. 9c), we find a general similarity in values. However, we also find substantial differences in observed and predicted calibration factors, suggesting the presence of other factors contributing to calibration factor variability. As suggested by Fig. 8, systematic variation in HF sensor heights could play an important role here. Other possible factors affecting calibration variability include: intensity of the emitted laser beam, clarity of the lens for the received laser beam, shape and orientation of particles passing through the beam, and optical properties of particles (Barchyn et al., 2014a). In general, though, the adoption of our calibration technique avoids the need to ascribe specific explanations to variations in HF sensor sensitivities through time.

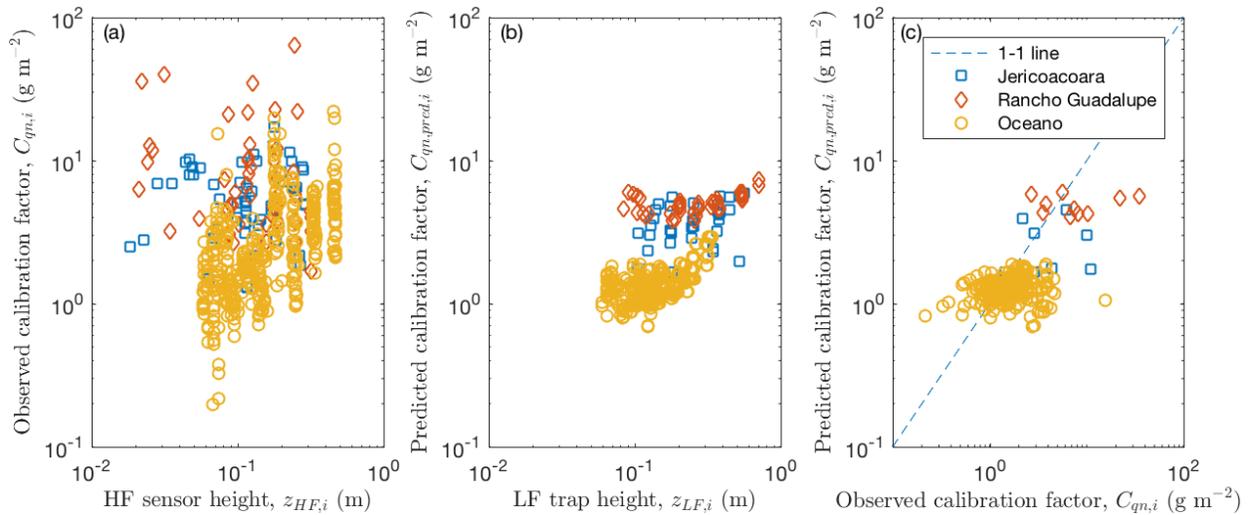

**Figure 9**. (a) Observed calibration factor $C_{qn,i}$ versus HF sensor height $z_{HF,i}$ (Eq. 14). (b) Variation in expected calibration factor $C_{qn,pred,i}$ (Eq. 15) versus LF trap height $z_{LF,i}$. (c) Direct comparison of observed and predicted calibration factors. Due to differences in trap heights

versus sensor heights, only a limited number of data points (with similar heights) are included in this plot.

---

We also evaluated the possibility that HF sensor saturation (i.e., reaching a maximum count rate), could affect the calibration coefficients. To do so, we examined all calibration factors $C_{qn,i}$ versus the number counts $n_{cal,HF,i}$ for all HF sensors $i$ for all calibration intervals (Fig. S4). Were saturation to occur, we would expect an increase in $C_{qn,i}$ for large $n_{cal,HF,i}$, corresponding to undercounting of particles. We do indeed note a slight but significant increase in $C_{qn,i}$ with $n_{cal,HF,i}$ at Jericoacoara and Oceano, but this increase only accounts for a small fraction of the variability in $C_{qn,i}$. In fact, for large $n_{cal,HF,i}$ ($> 100$ s$^{-1}$), there appear to be a few cases of unusually small $C_{qn,i}$, indicating either an overcounting of particles or an underprediction of the saltation flux during the calibration interval. Notably, these cases are all associated with small values of $z_{HF,i}$, suggesting the possibility that these HF sensors are embedded in a near-surface region in which saltation flux is enhanced beyond the exponential expectation (Namikas, 2003; Bauer and Davidson-Arnott, 2014). Thus, while there may be some bias in calibration coefficients related to variation in the saltation flux, such bias is unlikely related to sensor saturation.

### 4.3. Subsampling high-frequency (HF) saltation flux profiles
In this section, we describe methods for calibrating and combining saltation number counts from HF sensors to obtain height-specific horizontal saltation flux profiles $q_i(z_i)$ over arbitrary subsampling time intervals $\Delta t$. Through fits to such profiles, we can estimate total saltation fluxes $Q$, saltation layer heights $z_q$, and other profile parameters, as we will describe in Sec. 4.4 below. However, in preparing for this profile fitting, we sometimes faced issues of repeated HF sensor heights or zero flux values within profiles, potentially hindering the fitting process. Therefore, after first describing methods for obtaining individual values for calibrated height-specific fluxes $q_i$, we detail the occurrence of, and treatments for, repeated sensor heights and zero flux values in saltation flux profiles.

---

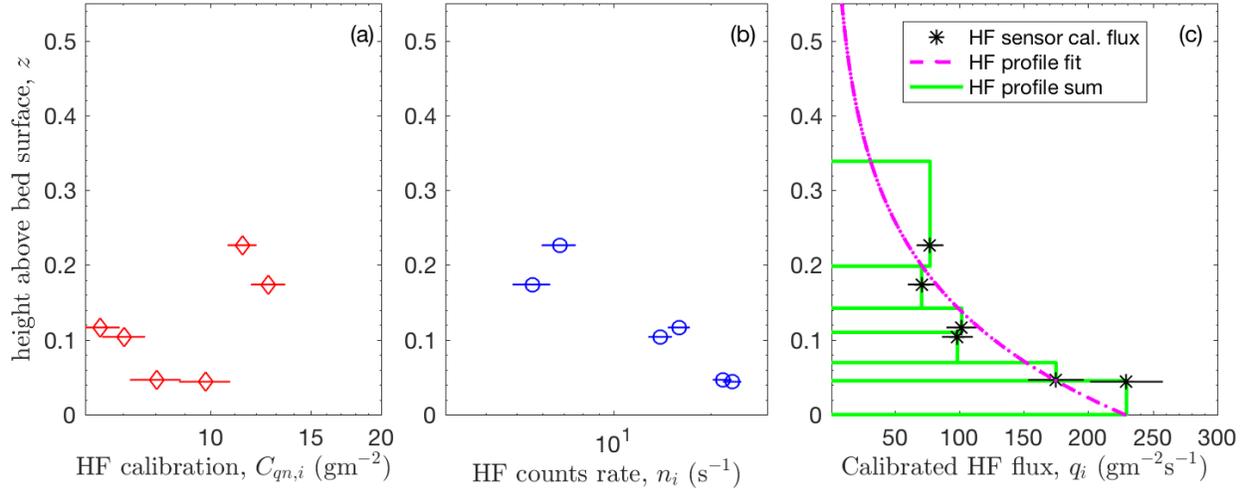

**Figure 10.** Demonstration of the three primary steps in the method for computing saltation flux for the profile of HF sensors for a $\Delta t = 10$s subsampling interval, as laid out in detail in Sec. 4.3. The data are from Jericoacoara on 14 November 2014, from 15:26:00 to 15:26:10 local time. (a) First, we obtain the calibration factors $C_{qn,i}$ from the corresponding LF trap time interval (i.e., Fig. 6c). (b) Second, we compute the HF sensor count rates $n_i$ for the $\Delta t = 10$s subsampling interval (blue circles). (c) Third, we compute calibrated saltation fluxes $q_i$ over $\Delta t$ as the product of $C_{qn,i}$ and $n_i$ by Eq. 10 (black asterisks). To estimate the total saltation flux $Q$, we can either apply the profile fitting method (dashed magenta line – Sec. 4.4.1) or the summation method (green bars – Sec. 4.4.2) for the profile of $q_i$.

### 4.3.1. Calculation of height-specific saltation fluxes for HF sensors

Given a subsampling time interval $\Delta t$, we compute count rates $n_i$ for each HF sensor $i$ as the total measured particle counts divided by $\Delta t$. We then apply Eq. 10 to convert each $n_i$ into a calibrated height-specific horizontal saltation flux $q_i$, using the calibration factor $C_{qn,i}$ obtained for the concurrent LF time interval, as illustrated in Fig. 10. Methods for estimating the associated calibrated height-specific horizontal saltation flux uncertainty $\sigma_{q_i}$ are described in Supporting Information S1.8.

### 4.3.2. Combination of saltation fluxes for repeated HF sensor heights

For a subset of the measurements collected at our Jericoacoara field site, multiple HF sensors were situated at identical heights. We could have included each of these calibrated height-specific saltation fluxes $q_i$ separately in the vertical profile fitting, but doing so would have required the exclusion of certain HF measurements for which individual sensors registered zero flux, i.e. $q_i = 0$. Instead, we choose to combine all of the $q_i$ from the same height into a single value, which we do by computing an uncertainty-weighted average (Eq. 4.17 in Bevington and Robinson (2003)) of these values. When we combine multiple $q_i$ into a single value for a certain time interval, we also combine the associated uncertainties of these individual $q_i$ into a single mean uncertainty (Eq. 4.19 in Bevington and Robinson (2003)).

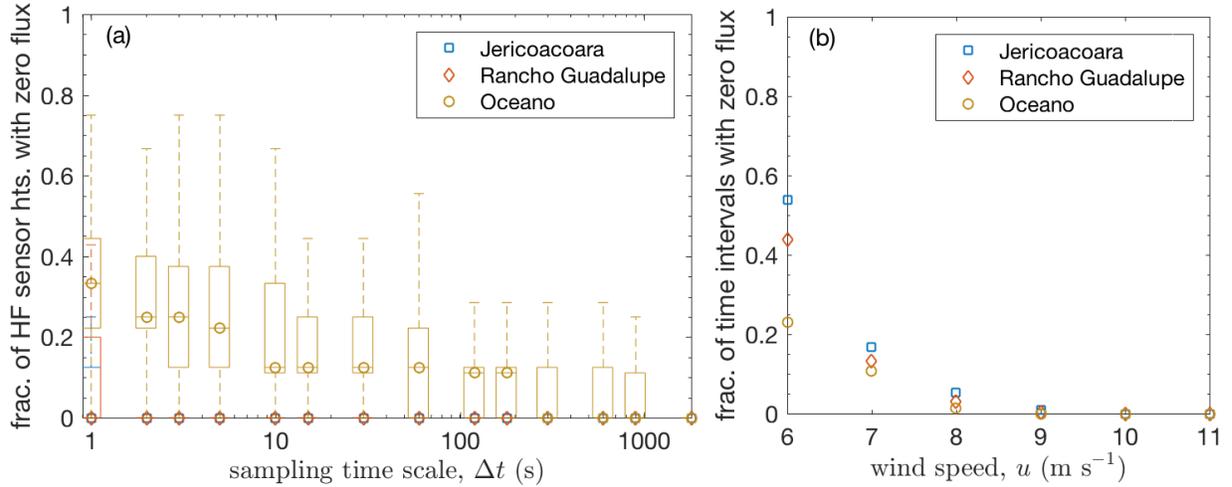

**Figure 11.** (a) Box plot describing the number of heights within the HF sensor profile for which height-specific saltation flux $q_i$ equals zero, as a function of time scale at each field site. Boxes indicate the range from 25$^{th}$ to 75$^{th}$ percentile values, and bars indicate the full range. For Jericoacoara and Rancho Guadalupe, most of the values are 0. (b) For a specific sensor height, $z_{HF,i}/z_q \approx 2$, and specific sampling interval $\Delta t = 5$ s, the fraction of time intervals for which $q_i$ equals zero, as a function of wind speed $u$.

### 4.3.3. Occurrence and treatment of zero values in vertical flux profiles

In certain instances, some, but not all of the HF sensors registered zero particle counts (i.e., $n_i = 0$) and thus zero calibrated flux ($q_i = 0$). However, performing exponential fits to $q_i(z_i)$ (see Sec. 4.4 below) requires first calculating $\log(q_i)$, which is not possible for $q_i = 0$. This is generally a minor issue over long averaging time intervals $\Delta t$, for which all sensors register particle counts. However, for small $\Delta t$, weak saltation, or sensors far above the surface, instances of $q_i = 0$ become increasingly common.

In Fig. 11, we examine the occurrence of zero values in saltation flux profiles. Fig. 11a shows how, at our Oceano field site, the incidence of $q_i = 0$ increases with decreasing sampling time scale $\Delta t$. Zero values are less of an issue at Jericoacoara and Rancho Guadalupe, where saltation tended to be stronger and HF sensors were closer to the surface. To control for these effects, we consider in Fig. 11b the occurrence of $q_i = 0$ for a fixed height $z_i$ and fixed sampling time scale $\Delta t$ as a function of wind speed, $u$. When controlling for these factors, HF sensors at all sites have a comparable rate of incidence of zero flux values.

Flux profiles with zeros are problematic. The simplest approach would be to simply exclude from analysis those time intervals for which at least one $q_i = 0$. However, doing so could preferentially exclude time intervals of weak saltation, thus reducing the data available for analysis and biasing the results from these analyses. An alternative approach is to exclude from analysis only those individual sensors with $q_i = 0$, but to continue with profile fitting to the remaining sensor heights with $q_i > 0$. Such selective inclusion of sensors is possible if there are at least 3 measurements available to perform exponential fits and compute uncertainties. However, such selective inclusion of certain sensor heights may bias vertical profile fits.

Fig. 12 considers these two approaches to the treatment of zero values in exponential profile fits at a variety of sampling time scales $\Delta t$. For each time interval with at least 3 HF heights for which $q_i > 0$, we perform an exponential profile fit using methods described in Supporting Information S1.3 to obtain a value for the saltation layer height $z_q$. Then, we follow two approaches to compute the average and uncertainty of $z_q$ values for a single site and $\Delta t$: first, we include only "full" profiles in this averaging (i.e., profiles for which all $q_i > 0$), and second, we include all profiles in this averaging (except for profiles with fewer than 3 $q_i > 0$, the minimum required for fitting). For the most part, there is no difference in these two methods for treating zero values, which occur very infrequently at Jericoacoara and Rancho Guadalupe and for large $\Delta t$ at Oceano. However, for small $\Delta t$ at Oceano, $z_q$ values from full profiles only are larger than values from all profiles and diverge from values computed at longer time scales. This may occur because full profiles are biased toward instances where the height-specific saltation flux near the top of the saltation profile is larger than would normally be expected. In particular, for a small $\Delta t$, there is a minimum detectable nonzero $q_i$ that may exceed the actual height-specific saltation flux at HF sensor heights near the top of the profile, which skews the $z_q$ of these full profile fits toward larger values. As $\Delta t$ then increases, the minimum detectable nonzero $q_i$ decreases until such biasing is no longer an issue. Given this biasing effect, we choose instead to follow the approach of fitting to all possible vertical profiles of saltation flux, and to simply exclude single $q_i = 0$ values from these profile fits when necessary.

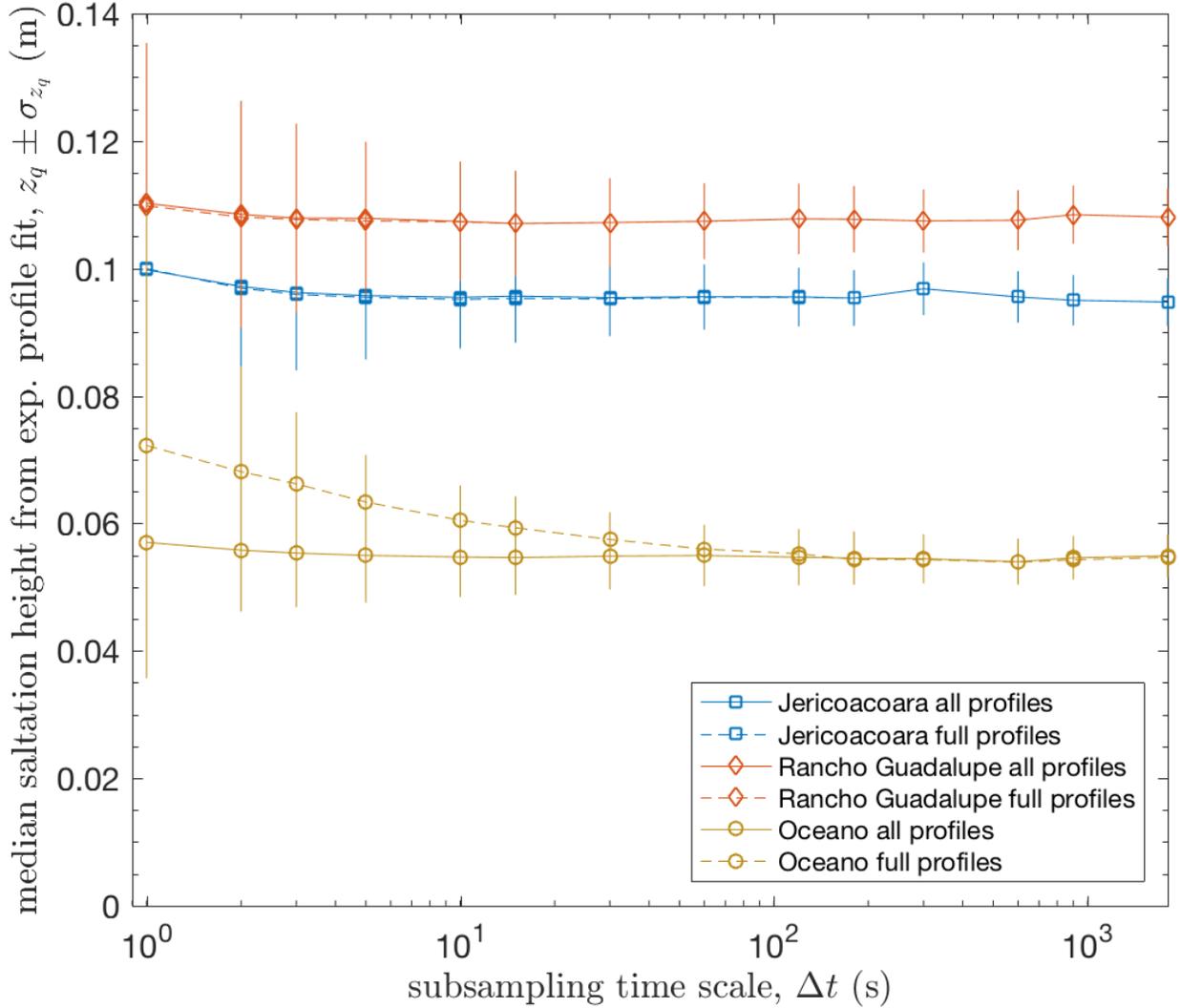

**Figure 12.** Median values of saltation layer height $z_q$ (symbols) and fitting uncertainty $\sigma_{z_q}$ (bars) for exponential fits to profiles of height-specific saltation flux $q_i$ from HF sensors, versus subsampling time scale $\Delta t$. Solid lines denote median $z_q$ computed from all flux profiles; dashed lines denote median $z_q$ computed only from "full" flux profiles for which $q_i > 0$ at all sensor heights.

### 4.4. Estimation of total (vertically-integrated) saltation flux

In this section, we describe methods for estimating the total saltation flux $Q$ by integrating over the vertical profiles of height-specific saltation flux $q_i(z_i)$ at arbitrary sampling time scales $\Delta t$. For these analyses, we follow the steps described in Sec. 4.3 above to obtain the $q_i(z_i)$ profiles. This includes using all flux profiles regardless of the existence of $q_i = 0$ within the profile, so long as there were at least 3 nonzero flux values for fitting the profile. Based on these profiles, we estimate the total flux as follows:

$$Q = \int_{z=0}^{z=\infty} q(z)\, dz. \tag{16}$$

Because we have only a limited number of $q_i(z_i)$ data points, calculation of $Q$ requires some approximations. In particular, based on our analysis in Sec. 4.1.4., we here assume an exponential profile for $q(z)$, thus:

$$Q = \int_{z=0}^{z=\infty} q_0 \exp\left(z/z_q\right) dz. \tag{17}$$

We consider two different methods for computing $Q$, as we describe below. In the first method, described in Sec. 4.4.1, we perform exponential profile fits to each $q_i(z_i)$ profile, then we use the resulting profile fit parameters $q_0$ and $z_q$ to estimate the total flux $Q$. In the second method, described in Sec. 4.4.2, we obtain the total flux $Q$ by computing a sum of all of the $q_i$, weighted according to their expected contributions to the exponential profile. Though requiring a greater number of assumptions to compute, the summation method allows $Q$ to be estimated at all time steps, regardless of the occurrence of $q_i = 0$ or other issues limiting an exponential profile fit.

### 4.4.1. Total flux estimation by exponential profile fit method

Given nonzero height-specific saltation fluxes $q_i$ at heights $z_i$, we perform exponential profile fits (Supporting Information S1.2) to obtain profile fit parameters $q_0$ and $z_q$. Calculating the integral in Eq. 17, we then obtain the total flux as:

$$Q_{fit} = q_0 z_q. \tag{18}$$

We calculate the associated uncertainty in $Q_{fit}$ through error propagation (see Supporting Information S1.9).

This method for obtaining $Q_{fit}$ assumes the ability to fit an exponential profile to the respective calibrated values of $q_i$ and $z_i$. When all of the $q_i$ equal 0, we assumed that $Q_{fit} = 0$. Otherwise, for time increments with some nonzero $q_i > 0$ but an insufficient number of these nonzero $q_i$ to perform an exponential fit (i.e., fewer than 3), the value for $Q_{fit}$ remained undefined. In cases of small $\Delta t$ or weak saltation, such undefined $Q_{fit}$ often constitute a substantial fraction of the time increments, hindering time series analyses for $Q_{fit}(t)$. To avoid this problem, the next section presents an alternative "weighted sum" approach to estimating the total, vertically-integrated, saltation flux.

### 4.4.2. Total flux estimation by weighted sum method

Over very short time increments $\Delta t$, performing the fit to obtain $q_0$ and $z_q$ for the site may be difficult if a large number of $q_i$ in the profile equal 0, or if the profile has not yet converged to its exponential form. In these cases, based on our observation that $z_q$ remains roughly invariant at each site, we can instead compute $Q$ as a weighted sum:

$$Q_{sum} = \sum_i \Delta Q_i, \tag{19}$$

where $\Delta Q_i$ is an incremental contribution to the total flux from $q_i$, weighted by the relative vertical coverage of $q_i$ in the profile. We compute $\Delta Q_i$ as:

$$\Delta Q_i = \int_{z_{i,bot}}^{z_{i,top}} q_{0,i} \exp\left(-\frac{z}{z_{q,LF}}\right) dz = q_{0,i} z_{q,LF} \left[\exp\left(-\frac{z_{i,bot}}{z_{q,LF}}\right) - \exp\left(-\frac{z_{i,top}}{z_{q,LF}}\right)\right], \quad (20)$$

where $z_{i,bot} = \sqrt{z_i z_{i-1}}$, and $z_{i-1}$ is the height of the HF sensor below $z_i$; if $z_i$ is the lowest, then $z_{i,bot} = 0$. $z_{i,top} = \sqrt{z_i z_{i+1}}$, and $z_{i+1}$ is the height of HF sensor above $z_i$; if $z_i$ is the highest, then $z_{i,top} = \infty$. $q_{0,i}$ is an equivalent value for $q_0$ determined based on $q_i$, as:

$$q_i = q_{0,i} \exp\left(-\frac{z_i}{z_{q,LF}}\right); \quad (21)$$

i.e., $q_{0,i}$ is the expected value of $q_0$ given an exponential profile with $q_i(z_i)$ and $z_{q,LF}$. In Eqs. 20 and 21, we use a saltation layer height equal to $z_{q,LF}$, i.e., the value obtained from the LF trap measurements prior to the calibration of the HF sensors. We calculate the uncertainty in $Q$ for the summation method through error propagation (see Supporting Information S1.10).

This summation method for obtaining $Q$ has the distinct advantage that it does not require convergence of the $q_i(z_i)$ profile to an exponential form, which can be an issue for fits applied to $q_i(z_i)$ at very small sampling time scales $\Delta t$. However, calculation of $Q$ by the summation method must assume a constant saltation layer height $z_{q,LF}$, whereas calculation of $Q$ by the exponential fitting method accounts for changes in $z_q$ over time. This limitation of the summation method is made acceptable by the fact that $z_q$ appears to remain roughly constant regardless of wind conditions and calculation time scale (Fig. 12 and Martin and Kok, 2017a).

### 4.4.3. Comparison of fitting and summation methods for estimating total saltation flux

In Fig. 13a we compare total saltation fluxes computed by the two methods, $Q_{fit}$ and $Q_{sum}$, as a function of sampling time scale $\Delta t$. We find that, whereas $Q_{sum}$ remains constant with $\Delta t$, $Q_{fit}$ varies with $\Delta t$ up to $\approx$ 30s. For $\Delta t \gtrsim$ 30s, there is negligible difference between $Q_{fit}$ and $Q_{sum}$. Uncertainties in $Q_{fit}$ and $Q_{sum}$ display trends similar to the flux values themselves (Fig. 13b). This divergence of $Q_{fit}$ from $Q_{sum}$ (and the associated uncertainties for these values) as $\Delta t$ becomes small suggests increasing error due to exponential fitting to limited data at very short time scales. These findings indicate that the $Q_{fit}$ method is reliable only for $\Delta t$ of at least 30s.

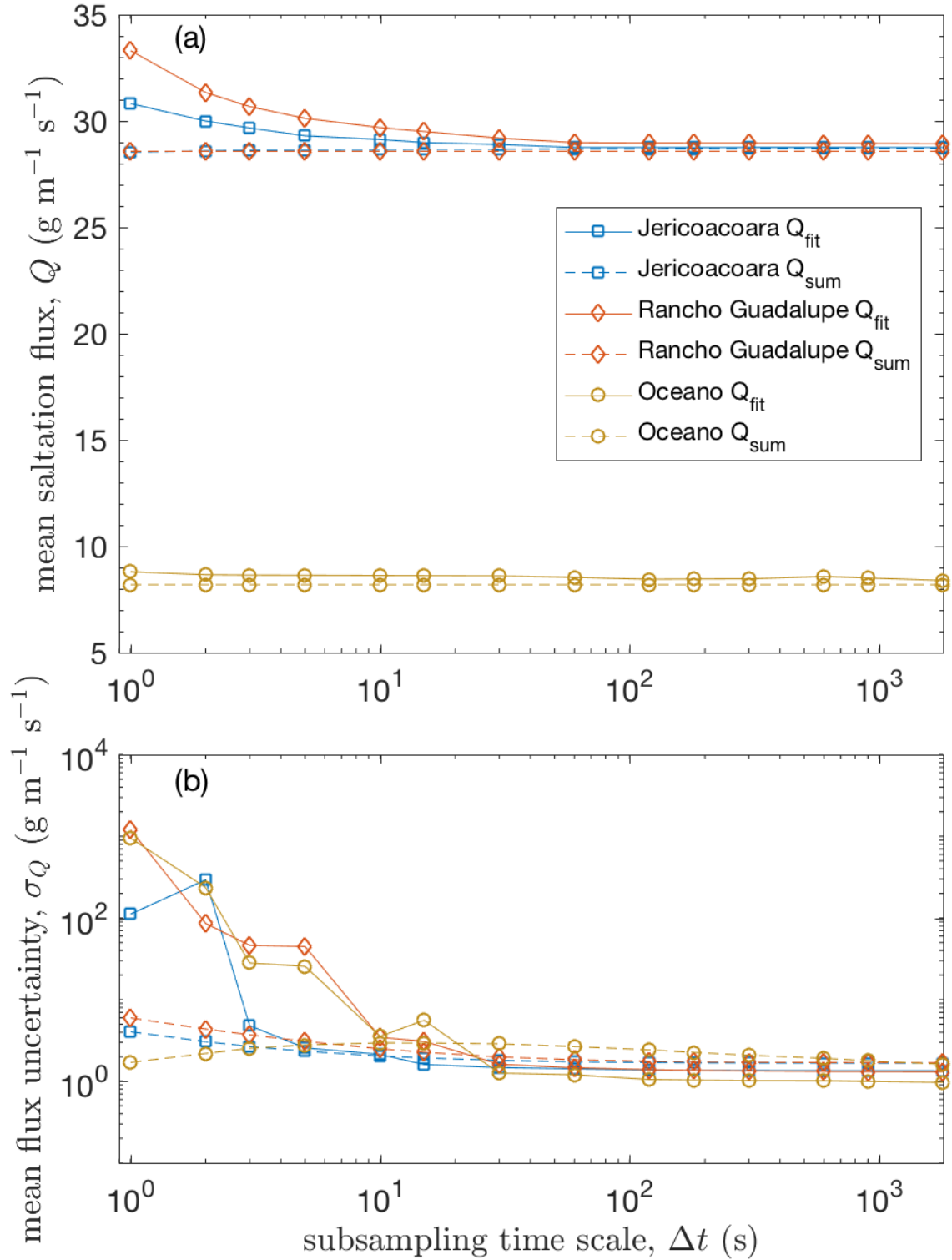

**Figure 13.** (a) Mean values for total saltation flux estimated by exponential profile fitting method $Q_{fit}$ (Eq. 18) versus summation method $Q_{sum}$ (Eq. 19). (b) Mean of uncertainty in exponential fit flux $\sigma_{Q_{fit}}$ (Eq. 32) and summation flux $\sigma_{Q_{sum}}$ (Eq. S35).

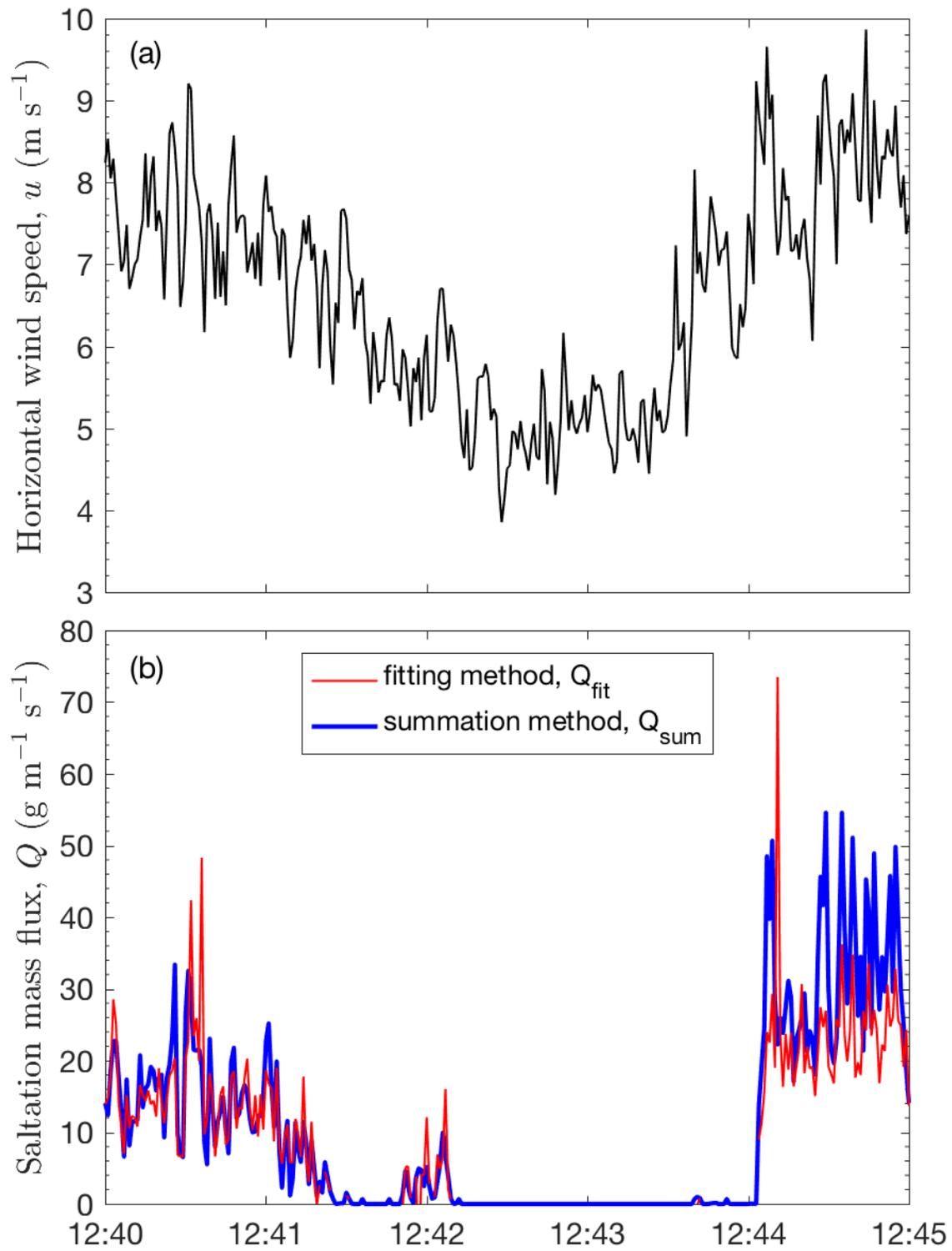

**Figure 14.** Sample time series of (a) streamwise wind speed $u$, and (b) total saltation mass flux $Q$ by fitting (Eq. 18) and summation (Eq. 19) methods calculated over $\Delta t = 1$ second

increments. The data are from Oceano on 16 May 2015, from 12:40 to 12:45 local time. Time increments with missing red curve indicate undefined values for $Q$ by the fitting method.

---

### 4.4.4. Illustration of calibrated high-frequency saltation flux time series

Though this paper is mostly focused on the methodology for characterizing high-frequency saltation flux, it is illustrative to examine how measured variations of saltation flux are related to measured variations of wind speed. As a basis for making the comparison between saltation and wind, we briefly describe here our methods for measuring wind speed properties, and we refer interested readers to examine these details further in Martin and Kok (2017a). For all wind calculations, we first subdivided raw wind data into 30-minute time intervals. For each of these 30-minute intervals, we performed a streamline correction to align the measured wind values with the mean streamwise wind $\bar{u}$, such that mean transverse $\bar{v}$ and vertical $\bar{w}$ winds equaled zero over each interval. Here, we consider only the streamline-corrected streamwise wind speed $u$.

A sample time series illustrating fluctuations in measured wind speed and saltation flux is shown in Fig. 14. This time series demonstrates the typical variability of streamwise winds $u$, which give rise to variations of total saltation flux $Q$. To first order, both the fitting and summation methods produce similar time series of $Q$. However, the fluctuations of $Q_{sum}$, which are not subject to noisy variations in the profile fitting process, tend to display smaller fluctuations than the values of $Q_{fit}$. Fig. 14 also shows that a small number of time increments only produce a value for $Q_{sum}$ but not for $Q_{fit}$. These gaps in $Q_{fit}$ occur when the exponential fitting fails, either due to a large number of zero $q_i$ values in the profile or a fitting to the remaining nonzero $q_i$ that fail to produce a real number for $Q$.

## 5. Discussion

We have presented a new methodology for determining the high-frequency variability of saltation flux under natural field conditions. We have shown that calibrated, height-specific saltation fluxes can be obtained at arbitrary time scales through the deployment of low-frequency (LF) saltation traps and high-frequency (HF) saltation sensors. We have further addressed high-frequency profile fitting issues by offering a summation method for the total saltation flux that is unaffected by the occurrence of poor profile fits at short time scales. This methodology, therefore, offers a novel way to directly compare high-frequency variations of wind speed and the total saltation flux, which can inform advances in understanding the two-way interactions between saltation transport and driving turbulent winds.

In this Discussion section, we compare our work with past attempts to characterize high-frequency variability of saltation flux, examining in particular how the new methodology addresses some, but not all, of the limitations of past work. Based on these considerations, we offer recommendations for future work to measure high-frequency variability of total saltation flux. We also explore some opportunities for addressing outstanding problems in aeolian research offered by our new methodology.

### 5.1. Comparison to previous high-frequency (HF) saltation characterization

The instruments used in our field deployments are not new to the study of aeolian saltation. The innovation of our work lies in the way we have combined these instruments, as the coupled use of LF traps and HF sensors in studies of aeolian saltation is rare (Martin et al., 2013). Past field-based studies examining high-frequency variability of aeolian saltation flux have typically only considered relative, and not absolute, variations of flux. When attempts have been made to convert HF number counts to absolute saltation fluxes, these studies have typically relied on theoretical conversion factors (Bauer and Davidson-Arnott, 2014), or used pre-determined empirical calibration factors (Nield et al., 2017). In contrast, our methodology employs a field-based calibration of HF sensor counts, which is directly tied to the specific conditions of the instruments and the field site. In the following subsections, we further consider the advancements of our methodology over past work, after which we describe some remaining limitations.

### 5.1.1. Advances over past work

One important element of our field-based empirical calibration procedures is that they allow for changes in the calibration factor through time, thus accommodating the issue of HF sensor drift observed in this and past studies (e.g., Bauer et al., 2012). These sensor drift issues appear to be most pronounced for optical sensors, such as the Wenglors used here (Hugenholtz and Barchyn, 2011), and for acoustic sensors, like the Miniphone (Ellis et al., 2009b). Sensor drift appears to be less of an issue for the generally more robust (but less sensitive) piezoelectric impact sensors (Hugenholtz and Barchyn, 2011); however, such sensors can have strong momentum / sampling area dependencies. When sensor drift occurs, it not only affects efforts to calibrate HF sensor counts to absolute saltation fluxes, but it also hinders comparisons of relative particle counts across sensors (Baas, 2008). Whether analyzing absolute flux or performing relative flux comparisons, our methodology therefore offers a potential improvement for the analysis of HF saltation time series from drift-prone sensors.

In addition to advancing methods for calculating HF height-specific saltation fluxes, we have also provided here new methods for obtaining HF time series of the total (vertically-integrated) saltation flux through a summation method that is unaffected by profile fit issues that arise at short time scales. In contrast, previous HF time series of total saltation flux typically depended only on saltation measured at a single height (e.g., Baas and Sherman, 2005). Based on the assumption, supported by observations, that the saltation layer is insensitive to changes in observational time scale (Fig. 12) or changes in wind conditions (Martin and Kok, 2017a), our summation method provides continuous time series of total saltation flux regardless of the intensity of saltation or the time scale of HF time series interval averaging.

### 5.1.2. Remaining limitations

Though useful for exploring many aspects of saltation mechanics, our methodology for characterizing high-frequency saltation flux depends on several assumptions, which limit its applicability in certain cases. These assumptions include the accuracy of LF trap measurements, negligible HF sensor drift within calibration intervals, spatial homogeneity within the deployment domain, and constant flux profile shapes during averaging intervals.

Our high-frequency flux calculation methodology relies on the assumption that saltation traps provide accurate and consistent measurements of saltation flux through time. However, studies have indicated the possibility that saltation trap efficiency may vary with airflow conditions and saltation intensities (Goossens et al., 2000; Sherman et al., 1998). Such biases could produce height-dependent relative errors in the LF profiles for $q(z)$, and thus systematic errors in the calibration factors used to obtain HF fluxes. However, these problems are not unique to the calibration of HF measurements, as they will equally affect LF saltation studies.

Our methodology for HF sensor calibration through comparison to LF trap profiles improves upon the method of Barchyn et al. (2014a), by not requiring extensive grain size measurements and limiting the need for multiple sensors at each height. Instead of requiring these additional measurements, the quality control in our methodology is achieved by examining the saltation profiles themselves. Such methodology thereby simplifies instrumentation requirements, but it also requires certain assumptions, like the existence of an exponential saltation profile, which may break down at small timescales, and a lack of variation in sensor performance at time scales that are less than the collection interval for LF traps measurements. To accommodate this latter assumption, we exclude HF sensors from our analysis if their particle count time series show anomalous changes in particle count rates compared to other sensors. However, we exclude only those sensors displaying the most egregious calibration issues (e.g., abrupt sensor failure in the middle of a calibration interval), leaving the possibility that we miss more minor calibration drift in certain sensors. Detecting and eliminating such HF sensors from analysis is therefore essential to the accuracy of these methods. In the future, increasing the number of sensors in a deployment may help identify anomalous sensor drift, presuming the mechanisms behind sensor drift are not affecting all sensors in a deployment (e.g., very dusty conditions changing sensitivity of Wenglor sensors).

In performing our calibration, we also assume spatial homogeneity within the footprint of HF and LF flux measurements. By making this assumption, we are able to relate the time-averaged LF trap vertical profile of saltation flux to the HF sensor vertical profile of particle counts. However, it is possible that, for spatially-separated LF and HF measurements, these profiles are not in fact measuring the same saltation flux. Indeed, at Oceano, where LF vertical profiles were separated in the spanwise direction by ~10 meters, we did observe significant profile differences (Supporting Fig. S3), possibly reflective of different surface or wind conditions. Because our HF measurements were situated at the halfway point between these two LF profiles, we use the average of the two LF profiles for calibration of the HF particle counts. This spanwise separation of LF profiles is likely reflected in the larger uncertainty of HF calibration factors at the Oceano site. Due to the assumption of spatial homogeneity, our saltation calibration methods are also unable to resolve spatial variability of saltation flux at length-scales shorter than the spatial extent of combined LF and HF instrument setups. Thus, to determine spatial variability related to vegetation (e.g., Stockton and Gillette, 1990; Davidson-Arnott et al., 2012; Barrineau and Ellis, 2013; Chapman et al., 2013), or dune topography (e.g., Bauer et al., 2012, 2015; Hoonhout and de Vries, 2017), multiple sets of co-located LF and HF instruments across the spatial gradient of interest would be required.

We also assume that vertical profiles of saltation flux follow an exponential form, with approximately constant saltation layer heights. We invoke the exponential profile at three points in our analysis: first, to fit the LF saltation flux profile; second, to predict time-integrated saltation fluxes at the heights of the Wenglors; and third, to fit the subsampled vertical profiles of HF saltation flux. Furthermore, we assume constant saltation layer height for these exponential profiles when computing uncertainty in exponential profile fits and when weighting the relative contributions of individual HF sensors for computing the total saltation flux by the summation method. Our observations of both LF (Figs. 5 and 6) and HF (Fig. 10) saltation fluxes seem to indicate that use of exponential profiles with constant saltation heights is reasonable, but other observations indicate variations of saltation profile shape through time (Bauer and Davidson-Arnott, 2014). This variation appears to occur mostly near the surface, i.e., $z < 2$ cm (e.g., Namikas, 2003; Bauer and Davidson-Arnott, 2014), where reptation and creep contribute significantly to the total sand flux. Due to the operational failure of our sensors so close to the surface, we were unable to obtain reliable HF saltation measurements in this zone, for which past measurements suggest that height-specific fluxes are larger than expected from the exponential profile (Namikas, 2003; Bauer and Davidson-Arnott, 2014). By excluding near-surface sensors, it is therefore possible that estimates of total saltation flux $Q$, by both the fitting and summation methods, may be underestimates of the true saltation flux. Further work is needed to develop robust HF saltation flux sensing methods close to the bed surface.

## 5.2. Recommendations for field work to characterize high-frequency (HF) saltation variability

Based on the innovations and limitations of our methodology for characterizing HF saltation flux variability, we now present recommendations for future field deployments to obtain HF saltation flux measurements. These recommendations include selection of instruments, strategies for deploying these instruments, and methods for processing LF and HF saltation data. Our recommendations are not meant to provide an authoritative endorsement of specific saltation traps and sensors. Rather, we hope to convey the main criteria for selecting, arranging, and utilizing saltation sensors for future field deployments. An overview of our instrumentation recommendations is provided in Table 2.

**Table 2: Instrumentation for field deployments to characterize high-frequency (HF) fluctuations of saltation mass flux**

| Instrument type | Instrument options | Points to consider |
| --- | --- | --- |
| Low-frequency (LF) saltation traps | BSNEs, MWACs, electronic traps, custom passive traps | Consistency of measurements, disruption of airflow, spatial footprint, required collection interval |
| High-frequency (HF) saltation sensors | Optical sensors (Wenglor, Nikolich), piezoelectric impact sensors (Sensit, Safire), acoustic impact sensors (Saltiphone, Miniphone), other techniques (e.g., videography) | Consistency, sensitivity, sampling interval, reliability and drift of measurements |
| Distance sensors | Laser sensors (Sick), acoustic sensors (Senix) | Accuracy, sampling area |
| Dataloggers | Campbell Scientific, Hobo | Sufficient ports and processing speed for data streams, synchronization, ruggedness, sampling frequency |

### 5.2.1. Low-frequency (LF) instrument selection

In our methodology for characterizing high-frequency (HF) saltation fluctuations, the role of the LF saltation traps is to provide vertical profiles of saltation flux in absolute mass flux units. Therefore, a key concern is that the LF traps provide reliable mass fluxes over the sampling intervals of interest. The reliability of LF traps was evaluated by Goossens et al (2000). They found that, among the 4 most commonly used saltation traps at that time, Modified Wilson and Cook (MWAC) traps provide the most consistent absolute measurements of saltation flux across a range of grain sizes and wind speeds, and that Big Springs Number Eight (BSNE) traps also provide reasonably consistent saltation fluxes, with high collection efficiency for all but the smallest (~132 um) particle size classes. In addition to the saltation traps described by Goossens et al (2000), a variety of other saltation traps have been developed in recent years, providing certain advantages in cost and portability (e.g., Namikas, 2002; Ridge et al., 2011; Sherman et al., 2014; Swann and Sherman, 2013; Hilton et al., 2017). A key consideration in deploying such traps is to ensure, through wind-tunnel or field tests, that the saltation fluxes measured by such traps are consistent with the measurements of known traps (e.g., Ridge et al., 2011). So long as the data are available from a study, and the traps can be demonstrated to be consistent with a known trap design, field study data will have longevity and the potential for future reuse. Without this, the data are only internally relative and may be of limited utility outside of the individual study (Barchyn et al., 2011).

In addition to LF saltation trap consistency, the spatial footprint of LF traps is also important. An issue with the BSNE traps used in our deployment is their large vertical range, which requires us to (1) estimate the representative height of each trap (Sec. 4.1) and (2) spatially separate the traps in the spanwise direction to provide higher resolution on the vertical profile (Sec. 2). Had we used traps with smaller cross-sectional areas, such as MWACs, we would have been able to achieve a closer vertical spacing of traps over a smaller areal footprint. However, by capturing smaller amounts of sand, use of such smaller traps would have possibly also introduced statistical uncertainty in saltation flux calculations, especially for time intervals when saltation is weak and the small error introduced by handling and weighing the sand becomes large relative to the volume of sand collected.

### 5.2.2. High-frequency instrument selection

In addition to practical concerns (cost, instrument degradation, deployment scheme), there are two primary technical considerations when choosing HF saltation sensors: consistency and sensitivity. For consistency, we desire that the sensor efficiency (i.e., particle counts per flux) remains roughly constant regardless of saltation intensity; otherwise, the empirical calibration we describe here is not feasible (though more careful, lab-based calibration could still be possible). Sensit piezoelectric impact sensors appear to provide the most consistent saltation response for a variety of weather conditions (Hugenholtz and Barchyn, 2011), though they are also the costliest and most insensitive HF sensors (Stout and Zobeck, 1997). In contrast, Safire piezoelectric impact sensors, though capable of sampling at 20 Hz or higher frequency, are subject to strong inconsistencies among instruments and over time (Baas, 2008). In particular, all circular piezoelectric sensors have an important momentum and sampling area bias that has relatively unquantified effects on the measurement efficacy (Barchyn and Hugenholtz, 2010). As for Saltiphone acoustic impact sensors, wind tunnel observations by Sterk et al. (1998) indicated consistent sensor mass flux responses in comparison to traps, whereas Goossens et al. (2000) showed that these sensors provide inconsistent results with changing wind speed, especially for coarser grain-size fractions. Wenglor optical counters appear to provide the most consistent particle counts over a reasonably wide range of saltation intensities (Hugenholtz and Barchyn, 2011), and, in contrast to past suggestion (e.g., Sherman et al., 2011), optical sensor saturation appears not to be a problem (Sec. 4.2). Optical sensors may also show a degradation in consistency due to scratching or buildup of dust on the lenses (Barchyn et al., 2014a). Recent work to develop a new optical sensor may address the issues of Wenglor sensor degradation (Etyemezian et al., 2017).

For sensitivity, we recommend sensors that are capable of detecting small particles and subtle variations of saltation flux. Such information is useful, in particular, for understanding the highest-frequency component of the saltation signal. However, for the purposes of determining saltation flux by our calibration method, sensitivity is less essential than consistency, because differences in HF sensor sensitivities will simply be reflected in differences of their calibration factors. In general, acoustic sensors appear to be more sensitive than piezoelectric sensors (Sherman et al., 2011), which display a range of sensitivities (Barchyn and Hugenholtz, 2010). Optical sensors appear to provide intermediate sensitivity (Hugenholtz and Barchyn, 2011; Leonard et al., 2011), though Wenglor sensors appear unable to detect small particles of <200 µm diameter (Duarte-Campos et al., 2017; Leonard et al., 2011). To overcome these issues of grain-size sensitivity bias, HF sensors should be developed in the future to directly detect airborne particle size distributions, such as through application of optical disdrometer techniques typically used for characterizing the size of rain droplets (e.g., Löffler-Mang and Joss, 2000).

### 5.2.3. Other measurements to include in saltation characterization deployments
In addition to LF and HF saltation measurements, we also included auxiliary measurements in our deployments to inform our calibration of HF saltation flux time series. At all of our field deployment sites, we mounted laser distance sensors with optical particle counters to detect variations of bed elevation (Sec. 3.2.3). Based on the known vertical displacement between distance sensors and saltation sensors, we could then calculate saltation sensor heights with respect to the bed surface (Fig. 7). In calculating saltation sensor heights (Sec. 4.2.2), we did not consider high-frequency variations of bed elevation associated with the passage of ripples, but instead considered the time-averaged bed elevation, representative of a spatial integration of upwind saltator sources. Distance sensors are thus useful primarily for detecting slower changes in the height of saltation sensors, such as those originating from settling of sensor towers or scour of the sand bed. We found the Sick DT35 laser distance sensors to provide relatively clean and accurate signals of bed elevation fluctuations (also useful for tracking ripple migration). In contrast, Senix acoustic distance sensors, tested at the Jericoacoara field site, provided much lower resolution and noisy bed elevation time series, due, in part, to the difficulty of electrically grounding these instruments in the dry, electrified environment of aeolian saltation.

Another essential instrument to include in a field deployment is a datalogger capable of recording the particle counts generated by HF sensors. We used Campbell CR1000 dataloggers, recording at 25 Hz, for this purpose. But other dataloggers, such as the Hobo (e.g., Davidson-Arnott et al., 2012), have also been commonly used in tandem with Wenglors and other saltation particle counters. When choosing dataloggers, it is important to ensure that they contain sufficient ports and processing speed to sustain the synchronous acquisition not just of HF saltation sensor data streams, but also data streams from distance sensors, anemometers, and other instruments. To record the many instruments in our field deployments, we synchronized the data records from multiple Campbell CR1000 dataloggers. Higher quality dataloggers have better internal clocks, which facilitate better time sync between various loggers.

### 5.2.4. Protocols for instrument layout
One key assumption underlying our calibration protocols is the comparability of LF and HF saltation flux measurements. A key consideration in laying out instruments, therefore, is to ensure, as much as possible, that measurements from traps and sensors are temporally and spatially coincident. Temporal coincidence can be ensured for LF traps by coordinating trap collection times as closely as possible; for HF sensors, by properly synchronizing datalogger pulse counting. Such synchronization of manually-recorded LF trap time intervals with automated HF time-stamping is primarily a logistical issue. At the beginning and end of each field day, field workers should check that all datalogger clocks and other timing devices used for recording time intervals are synchronized.

For spatial coincidence, the characteristic spatial and temporal variability of saltation must be considered. Saltation varies spatially over spanwise distances greater than ~10 cm with the passage of streamers on ~1 s timescales (Baas and Sherman, 2005). Ideally, each HF sensor would be exactly co-located with a calibrating LF trap to reduce this variation, and all sensors and traps would be commonly aligned (i.e., both rigid or both rotating). In practice, instruments need to be spatially separated and mounted to accommodate their size, reduce airflow interference among instruments, and fulfill other deployment needs (in our case, mounting HF sensors on a rigid array to accommodate distance sensor measurements of ripple migration). As a compromise, assuming spatial homogeneity of site conditions, measurements from broadly spatially separated LF traps (~meters) should be comparable over sufficient averaging times. However, if LF traps cover a broad spatial domain displaying heterogeneity in grain size or topography, instruments may no longer provide comparable measurements, as we observed in the spanwise-separated LF trap profiles at the Oceano field site. Furthermore, combining LF and HF sensors over a large spatial domain limits the possibilities for measuring spatial heterogeneity of saltation flux, such as associated with variability in dune topography (e.g., Bauer et al., 2012), vegetation (e.g., Chapman et al., 2013), or soil moisture (e.g., Arens, 1996). Together, these compromises suggest co-locating instrumentation to the greatest extent possible, while taking care to minimize airflow interference. If larger spatial extents are required, ensure homogeneity in conditions.

### 5.2.5. Selection of sampling time intervals
There are three time intervals associated with our method of characterizing high-frequency variation of saltation flux. First, there is the time interval for low-frequency (i.e., trap) measurements. This time interval should be long enough to ensure that all traps capture enough sand such that the errors associated with weighing the sand are small relative to sand mass. However, LF measurement intervals must be short enough that such collections can keep pace with possible changes in the calibration of the high-frequency (HF) sensors. Second, there is the sampling interval of HF sensors. This sampling interval will be determined, in part, by the technical capabilities of the saltation sensors and dataloggers used for the deployment. Ideally, the sampling interval will be as small as possible; but, in practice, the minimum sampling interval can be chosen according to the specific goals of the field deployment and the equipment available. Third, there is the subsampling interval $\Delta t$ for characterization of the saltation flux for the high-frequency sensors. In principle, it is possible to reduce $\Delta t$ to the sensor sampling interval; however, as we described in Sec. 4.3 above, fitting of saltation flux profiles becomes increasingly difficult for small $\Delta t$, especially when saltation is weak. Furthermore, if HF sensors are spatially separated, they may not provide comparable measures of saltation flux for $\Delta t <$ 0.1s due to the spatial variability of aeolian streamers.

### 5.2.6. Protocols for calibration

In Sec. 4.2, we provided a detailed methodology on how to calibrate relative HF sensor particle counts to absolute saltation mass fluxes. Though justified by some consideration of expected calibration factors resulting from spatial variation of airborne grain sizes (Fig. 8), our calibration scheme was empirical in nature and subject to broad unexplained variability (Fig. 7). We were able to perform such an empirical calibration because our deployments included a large number of both LF and HF saltation measurements in relative close proximity. Based on the utility of this calibration technique, we encourage other researchers to adopt similar instrumentation, instrument configurations, and calibration protocols in their field work. To follow our calibration protocols, field deployments to measure streamwise gradients in the high-frequency component of saltation flux time series, for example, will require the deployment of sets of HF and LF sensors at multiple locations, requiring careful planning and sufficient resources. To simplify these deployments, it may be possible to collocate a single LF trap and HF sensor at the same height at each location along the transect. In so doing, information about vertical saltation profiles will be lost; but, assuming the relative invariance of vertical profiles of saltation flux through time (Fig. 10), this may not be a major problem, especially if LF traps and HF sensors are designed to capture a large vertical range of the saltation profile.

### 5.2.7. Possible improvements for future work

Above, we provided suggestions for instrument selection, layout, sampling, and calibration protocols for obtaining high-frequency time series of saltation flux. Despite the potential improvements provided by our methodology in comparison to past work, challenges remain (see Sec. 5.1.2). Here, we offer several suggestions for future work to improve the quality of field-based studies of the high-frequency variability in saltation flux.

To address potential inconsistencies among LF saltation trap measurements, further work should be undertaken to directly compare the performance of the many saltation traps that have proliferated over the past few years (e.g., Namikas, 2002; Ridge et al., 2011; Swann and Sherman, 2013; Sherman et al., 2014; Hilton et al., 2017). In particular, traps to estimate not only airborne saltation fluxes but also the fluxes of reptating and creeping particles very close (<2 cm) to the sand surface (e.g., Swann and Sherman, 2013) require further evaluation. These trap designs should be standardized such that any calibration work can be applied widely.

Dealing with issues in HF sensor performance will require more than further testing, which has been quite thorough in recent years (Stout and Zobeck, 1997; Sterk et al., 1998; Goossens et al., 2000; Baas, 2004; Ellis et al., 2009b; Barchyn and Hugenholtz, 2010; Barchyn et al., 2014a; Duarte-Campos et al., 2017). HF sensors display common issues with calibration drift through time; such issues are clearly apparent in our field data (Fig. 8). By tuning relative HF sensor particle counts to absolute LF saltation trap fluxes, our calibration technique provides a way to address many of these sensor drift issues. However, application of our calibration method requires elaborate deployments of LF traps and HF sensors, and it does not address the underlying issues causing sensor drift. New sensor technologies may overcome these drift issues. For example, the recently developed optical "Nikolich" sensors (Etyemezian et al., 2017) directly record the time series of laser transmission used to detect particle counts, making it possible to perform post-processing to directly correct for long-term drift or other issues in sensor deployment. By providing more than just the particle counts, the records from such sensors could be carefully examined to build a more detailed understanding of how optical particle counters are affected by variations in particle size, sensor detection limits, laser beam properties, and ambient lighting conditions. For example, with such information, it would be possible to perform a more systematic analysis of the variation in calibration factor with particle size (Barchyn et al., 2014a; Duarte-Campos et al., 2017), thus instilling confidence in attempts to calibrate HF sensor counts without associated LF trap measurements (e.g., Bauer and Davidson-Arnott, 2014; Nield et al., 2017).

To reduce the effects of spatial instrument separation on the comparability of LF and HF saltation measurements in the calibration process, we suggest improvements to reduce the spatial footprint of saltation traps and sensors. Doing so would reduce airflow interference and allow the co-location of more measurements within a smaller spatial footprint. Further studies to examine how spanwise variability of saltation flux is related to instrument separation distance and measurement time scale could help to inform the needs of future saltation trap and sensor design. Recent work is also helping to inform understanding of how variable site characteristics affect variability in saltation flux (e.g., Webb et al., 2016a).

Differences in saltation measurement methods have inhibited inter-study comparisons of aeolian processes, leading for a call to adopt a common set of "standards" for the collection of aeolian field data (Barchyn et al., 2011). In response to this call, Webb et al. (2016b) have adopted a standard selection and configuration of instruments for their National Wind Erosion Research Network. Our methodology offers an alternative approach to inter-study comparisons that does not necessarily require such a standardized set of instruments. Instead, by providing a detailed set of protocols for combining LF trap and HF sensor data to characterize high-frequency variability in saltation flux driven by atmospheric turbulence, our methodology offers the possibility for future comparisons of data from across studies utilizing different combinations of instruments. Future work could evaluate the appropriateness of different traps and sensors in performing such inter-study comparisons.

### 5.3. Opportunities offered by high-frequency saltation datasets
Our calibrated HF field measurements of aeolian saltation offer unprecedented detail on saltation flux profiles and variability under a range of natural conditions. To our knowledge, the longest comparable field dataset was presented by Sherman et al (1998). However, the duration of the Sherman et al (1998) data set is only about 15 hours, compared to the 75 hours of active saltation that we measured, and the temporal resolution of their data (1 Hz) is substantially lower than ours (25 Hz). Baas and Sherman (2005) collected field data with comparable temporal resolution as ours, but their data span less than 1 hour of active saltation. Our field data therefore offer substantial opportunities to understand various aspects of saltation process mechanics. These data have already been used to investigate the saltation flux law (Martin and Kok, 2017a) and saltation thresholds (Martin and Kok, 2017b). Further analyses could reveal how interactions between atmospheric turbulence and the saltation layer structure affect variability in saltation flux (e.g., Baas and Sherman, 2005), the saltation saturation time scale (Pähtz et al., 2013), and momentum transport in the saltation layer (Sherman and Farrell, 2008; Li et al., 2010). Because the dust emission flux is roughly proportional to the sand flux (Shao et al., 1993; Marticorena and Bergametti, 1995; Kok et al., 2014), these data also offer opportunities for the development of more accurate dust emission models.

## 6. Conclusion

In this paper, we presented a new methodology for characterizing high-frequency fluctuations of aeolian saltation flux. Our methodology combines low-frequency (LF) saltation traps that provide absolute saltation fluxes and high-frequency (HF) sensors that detect relative changes of saltation fluxes on short time scales. The increased accuracy of high-frequency saltation flux estimates facilitated by this new approach is a prerequisite for understanding how saltation responds to variations of wind and associated turbulence structures. As such, the measurements and analysis workflow presented here could play an important role in developing more accurate aeolian transport models that move beyond the common assumption of steady-state time-averaged saltation dynamics.

**ACKNOWLEDGEMENTS.** This work was supported by U.S. National Science Foundation (NSF) Postdoctoral Fellowship EAR-1249918 to R.L.M. and NSF grants AGS-1358621 to J.F.K., AGS-1358593 to M.C., and a Core Fulbright U.S. Scholar Program award to J.T.E. Research was also sponsored by the Army Research Laboratory and was accomplished under Grant Number W911NF-15-1-0417. The views and conclusions contained in this document are those of the authors and should not be interpreted as representing the official policies, either expressed or implied, of the Army Research Laboratory or the U.S. Government. The U.S. Government is authorized to reproduce and distribute reprints for Government purposes notwithstanding any copyright notation herein. Jericoacoara fieldwork is registered with the Brazilian Ministry of the Environment (#46254-1 to J.T.E.). Oceano Dunes State Vehicular Recreation Area, Rancho Guadalupe Dunes Preserve, and Jericoacoara National Park provided essential site access and support. We thank Doug Jerolmack for lab access for grain-size analysis, and Paulo Sousa, Peter Li, Francis Turney, Arkayan Samaddar, and Livia Freire for field assistance. Data and scripts associated with the methodology presented here may be obtained by contacting the authors.

# Supporting information

"High-frequency measurements of aeolian saltation flux: Field-based methodology and applications"

**Overview**

In this supplementary document, we present statistical methods used in the analysis of saltation flux profiles (Sec. S1). We also provide a list of variable names used in the manuscript (Sec. S2).

## S1. Statistical methods

Here, we provide details on line and profile fitting procedures invoked repeatedly within the main text of this paper.

### S1.1. Uncertainty estimation and propagation

Given variables $v_1, v_2, \ldots v_i$ included in the calculation of a variable $s$, and assuming uncorrelated uncertainties $\sigma_{v_1}, \sigma_{v_2}, \ldots \sigma_{v_i}$ for the constituent variables, we compute the uncertainty $\sigma_s$ in $s$ by applying the error propagation formula (Eq. 3.14 in Bevington and Robinson, 2003):

$$\sigma_s = \sqrt{\sigma_{v_1}^2 \left(\frac{\partial s}{\partial v_1}\right)^2 + \sigma_{v_2}^2 \left(\frac{\partial s}{\partial v_2}\right)^2 + \cdots + \sigma_{v_i}^2 \left(\frac{\partial s}{\partial v_i}\right)^2}. \quad (S1)$$

### S1.2. Linear fitting procedure

Given independent variable data points $x_i$ and dependent variable data points $y_i$ with uncertainties $\sigma_{y_i}$, we compute linear fits to:

$$y_{fit,i} = a + bx_i, \quad (S2)$$

where $y_{fit,i}$ is the fitted value for the dependent variable corresponding to $x_i$, $a$ is the fitting intercept, and $b$ is the fitting slope. We compute fit values $a$ and $b$ and their respective uncertainties $\sigma_a$ and $\sigma_b$ based on Eqs. 6.12, 6.21, and 6.22 in Bevington and Robinson (2003) applied to the known $x_i$, $y_i$, and $\sigma_{y_i}$. We then determine uncertainty in these fit values based on the error propagation formula for correlated variables (Eq. 3.13 in Bevington and Robinson, 2003):

$$\sigma_{y_{fit,i}} = \sqrt{\sigma_a^2 \left(\frac{\partial y_{fit,i}}{\partial a}\right)^2 + \sigma_b^2 \left(\frac{\partial y_{fit,i}}{\partial b}\right)^2 + \sigma_{ab}^2 \frac{\partial y_{fit,i}}{\partial a}\frac{\partial y_{fit,i}}{\partial b}}, \quad (S3)$$

where we determine the fitting parameter covariance $\sigma_{ab}^2$ based on Eq. 7.23 in Bevington and Robinson (2003). Computing $\frac{\partial y_{fit,i}}{\partial a}$ and $\frac{\partial y_{fit,i}}{\partial b}$ for Eq. S2 and plugging these into Eq. S3 gives the uncertainty on linear fit values:

$$\sigma_{y_{fit,i}} = \sqrt{\sigma_a^2 + \sigma_b^2 x_i^2 + 2\sigma_{ab}^2 x_i}. \tag{S4}$$

### S1.3. Exponential profile fitting procedure

Based on known saltation fluxes $q_i$ at heights $z_i$ with associated uncertainties $\sigma_{q_i}$ and $\sigma_{z_i}$ respectively, we compute fits to exponential flux profiles of the form:

$$q_{exp,i} = q_0 \exp\left(-\frac{z_i}{z_q}\right), \tag{S5}$$

where $q_{exp,i}$ are exponentially-fitted flux values with uncertainties $\sigma_{q_{exp,i}}$, $q_0$ is the fitted flux scaling parameter with uncertainty $\sigma_{q_0}$, and $z_q$ is the fitted characteristic *e*-folding saltation layer height with uncertainty $\sigma_{z_q}$. To apply the linear fitting procedure described above in S1.2, we relate the linear fit variables in Eq. S2 to the exponential fit variables in Eq. S5 as follows:

$$x_i = z_i, \tag{S6}$$

$$y_i = \log(q_i), \tag{S7}$$

$$\sigma_{y_i} = \sqrt{\frac{\sigma_{q_i}^2}{q_i^2} + \frac{\sigma_{z_i}^2}{z_q^2}}, \tag{S8}$$

where we calculate Eq. S8 by assuming uncorrelated uncertainties for $\sigma_{q_i}$ and $\sigma_{z_i}$ and by applying the error propagation formula (Eq. S1) to Eqs. S5 and S7. Prior to calculating the $\sigma_{y_i}$ in Eq. S8, we do not know specific values for $z_q$; therefore, we use the site-averaged values for $z_q$, which remain roughly constant through time (Martin and Kok, 2017a), when performing this uncertainty propagation. (Alternatively, when performing profile fits for HF sensor saltation flux profiles, we use $z_q$ from the LF trap profile fit for the corresponding calibration time interval.)

Based on these computed values for $x_i$, $y_i$, and $\sigma_{y_i}$, we then perform the linear fit described by Eq. S2. We convert the resulting linear fit parameters to exponential fit parameters as:

$$q_0 = \exp(a), \tag{S9}$$

$$z_q = -1/b. \tag{S10}$$

Based on the linear fit parameter uncertainties ($\sigma_a, \sigma_b$) we estimate exponential fit parameter uncertainties by applying the uncorrelated error propagation formula (Eq. S1) to Eqs. S9 and S10:

$$\sigma_{q_0} = \sigma_a q_0, \tag{S11}$$

$$\sigma_{z_q} = \sigma_b z_q^2. \tag{S12}$$

We estimate exponential fit parameter covariance as:

$$\sigma^2_{q_0,z_q} = \sigma^2_{a,b} \frac{\partial q_0}{\partial a} \frac{\partial z_q}{\partial b} = \sigma^2_{a,b} q_0 z_q^2, \quad (S13)$$

where $\sigma^2_{a,b}$ is the linear fit covariance.

**S1.3. Power-law profile fitting procedure**

Based on known saltation fluxes $q_i$ at heights $z_i$ with associated uncertainties $\sigma_{q_i}$ and $\sigma_{z_i}$ respectively, we compute fits to power law flux profiles of the form:

$$q_{pwr,i} = q_p z_i^{-k_z}, \quad (S14)$$

where $q_{pwr,i}$ are fitted flux values with uncertainties $\sigma_{q_{pwr,i}}$, $q_p$ is the fitted flux scaling parameter with uncertainty $\sigma_{q_p}$, and $k_z$ is the fitted characteristic exponent for the fit with uncertainty $\sigma_{k_z}$. To apply the linear fitting procedure described above, we relate the linear fit variables in Eq. S2 to the power law fit variables in Eq. S14 as follows:

$$x_i = \log(z_i), \quad (S15)$$

$$y_i = \log(q_i), \quad (S16)$$

$$\sigma_{y_i} = \sqrt{\frac{\sigma^2_{q_i}}{q_i^2} + \frac{k_z^2 \sigma^2_{z_i}}{z_i^2}}, \quad (S17)$$

where we determine Eq. S15 by assuming uncorrelated uncertainties for $\sigma_{q_i}$ and $\sigma_{z_i}$ and by applying the error propagation formula (Eq. S1) to Eqs. S15 and S16. Because we do not know the value for $k_z$ prior to fitting, we assume that $k_z \approx 1$, thus

$$\sigma_{y_i} = \sqrt{\frac{\sigma^2_{q_i}}{q_i^2} + \frac{\sigma^2_{z_i}}{z_i^2}}, \quad (S18)$$

Based on these computed values for $x_i$, $y_i$, and $\sigma_{y_i}$, we then perform the linear fit described by Eq. S2. We convert the resulting linear fit parameters to power law fit parameters as:

$$q_p = \exp(a), \quad (S19)$$

$$k_z = -b. \quad (S20)$$

Based on the linear fit parameter uncertainties ($\sigma_a$, $\sigma_b$) we estimate power law fit parameter uncertainties by applying the uncorrelated error propagation formula (Eq. S1) to Eqs. S19 and S20:

$$\sigma_{q_p} = \sigma_a q_p, \quad (S21)$$

$$\sigma_{k_z} = \sigma_b. \quad (S22)$$

We estimate power law fit parameter covariance as:

$$\sigma^2_{q_p, k_z} = \sigma^2_{a,b} \frac{\partial q_p}{\partial a} \frac{\partial k_z}{\partial b} = \sigma^2_{a,b} q_p, \quad (S23)$$

where $\sigma^2_{a,b}$ is the linear fit covariance.

### S1.4. Uncertainty estimation for LF trap fluxes

To estimate the uncertainty $\sigma_{q_{LF,i}}$ in the height-specific horizontal saltation flux $q_{LF,i}$ for the *i*th LF saltation trap, we perform the following calculation through application of the error propagation formula (Eq. S1) to the flux calculation (Eq. 1 in main text):

$$\sigma_{q_{LF,i}} = q_{LF,i} \sqrt{\frac{\sigma^2_{m_{LF,i}}}{m^2_{LF,i}} + \frac{\sigma^2_{H_{LF}}}{H^2_{LF}} + \frac{\sigma^2_{W_{LF}}}{W^2_{LF}} + \frac{\sigma^2_{T_{LF}}}{T^2_{LF}}}. \quad (S24)$$

We estimate LF trap height uncertainty ($\sigma_{H_{LF}} = 0.1$ cm) and width uncertainty ($\sigma_{W_{LF}} = 0.1$ cm) based on measurement error in the size of the trap opening, and we estimate LF duration uncertainty ($\sigma_{T_{LF}} = 30$ s) based on the typical variability in the time required to cover and uncover all of the LF traps. The LF mass uncertainty $\sigma_{m_{LF,i}}$ included both relative uncertainty on trap collection efficiency ($\sigma_{trap} = 10\%$ for BSNEs, Goossens et al., 2000) and absolute uncertainty in the weight of the collected sand ($\sigma_{weight} = 0.1$ g) based on the accuracy of the weighing scale. We combine these mass uncertainties in quadrature as:

$$\sigma_{m_{LF,i}} = \sqrt{\sigma^2_{trap} m^2_{LF,i} + \sigma^2_{weight}}. \quad (S25)$$

### S1.5. Uncertainty estimation for combining LF trap profiles

In Sec. 4.1.5 of the main text, we described methods to combine exponential profile fit values for LF saltation traps separated by a large distance in the spanwise direction. Here, we describe calculations for the corresponding uncertainties of these combined values.

Denoting the separate LF profile fit values as $z_{q,LF,+}$ and $q_{0,LF,+}$ (with respective uncertainties $\sigma_{z_{q,LF,+}}$ and $\sigma_{q_{0,LF,+}}$) for the +$y$ LF profile fits and $z_{q,LF,-}$ and $q_{0,LF,-}$ (with respective uncertainties $\sigma_{z_{q,LF,-}}$ and $\sigma_{q_{0,LF,-}}$) for the −$y$ LF profile fits, we estimate uncertainties for combined fit values $z_{q,LF}$ (Eq. 8 in main text) and $q_{0,LF}$ (Eq. 9 in main text) through application of the error propagation formula (Eq. S1):

$$\sigma_{z_q,LF} = \frac{1}{2}\sqrt{\sigma_{z_q,LF,+}^2 + \sigma_{z_q,LF,-}^2}, \tag{S26}$$

$$\sigma_{q_0,LF} = \frac{1}{2}\sqrt{\sigma_{q_0,LF,+}^2 \frac{q_{0,LF,-}}{q_{0,LF,+}} + \sigma_{q_0,LF,-}^2 \frac{q_{0,LF,+}}{q_{0,LF,-}}}. \tag{S27}$$

We then approximate the combined covariance $\sigma_{q_0,z_q,LF}^2$ as the geometric average of the two covariances, $\sigma_{q_0,z_q,LF,+}^2$ and $\sigma_{q_0,z_q,LF,-}^2$, for the individual spanwise-separated profiles.

### S1.6. Uncertainty estimation for predicted HF saltation fluxes

To estimate uncertainty of the predicted height-specific saltation flux $q_{pred,HF,i}$ at HF sensor height $z_{HF,i}$, we apply the error propagation formula for correlated uncertainties (Eq. S3) to Eq. 11 in the main text:

$$\sigma_{q_{pred,HF,i}} = q_{pred,HF,i}\sqrt{\frac{\sigma_{q_0,LF}^2}{q_{0,LF}^2} + \frac{\sigma_{z_q,LF}^2 z_{HF,i}^2}{z_{q,LF}^4} + 2\sigma_{q_0,LF,z_q,LF}^2 \frac{z_{HF,i}}{q_{0,LF} z_{q,LF}^2} + \frac{\sigma_{z_{HF,i}}^2}{z_{q,LF,i}^2}}. \tag{S28}$$

where $\sigma_{q_0,LF}$ and $\sigma_{z_q,LF}$ are the LF fitting parameter uncertainties, $\sigma_{q_0,LF,z_q,LF}^2$ is the covariance of these uncertainties, and $\sigma_{z_{HF,i}}$ is the uncertainty in HF sensor height. We assume that $\sigma_{z_{HF,i}} = \sigma_{z_{rel,i}} = 1$ mm, i.e., the error in measuring the height of the Wenglor and distance sensor on their mounting stand. We neglect uncertainty in the bed height $z_{dist}$ measured by the distance sensor, because such uncertainty is correlated across all sensors heights.

### S1.7. Uncertainty estimation for HF sensor saltation flux calibration factors

To estimate uncertainty of the saltation flux calibration factor $\sigma_{C_{qn,i}}$ for each HF sensor *i*, we propagate the uncertainty in $\sigma_{q_{pred,HF,i}}$ through application of Eq. S1 to Eq. 14 in the main text:

$$\sigma_{C_{qn,i}} = \frac{\sigma_{q_{pred,HF,i}}}{n_{cal,i}}. \tag{S29}$$

In computing $\sigma_{C_{qn,i}}$, we neglect uncertainty of the particle count rate $n_{cal,i}$, which we will consider below (Sec. S1.8) when applying the calibration factor to compute uncertainty in calibrated HF sensor saltation fluxes for subsampled time intervals.

### S1.8. Uncertainty estimation for calibrated subsampled HF sensor saltation flux

To estimate uncertainty of the calibrated saltation flux $q_i$ for HF sensor *i* computed over a subsampling time interval $\Delta t$, we combine the contributions of uncertainty from the calibration factor $\sigma_{C_{qn,i}}$ and the number counts $\sigma_{n_i}$ through application of Eq. S1 to Eq. 10 in the main text:

$$\sigma_{q_i} = \sqrt{\sigma_{C_{qn,i}}^2 n_i^2 + \sigma_{n_i}^2 C_{qn,i}^2}. \tag{S30}$$

Here, we use $\sigma_{C_{qn,i}}$ from Eq. S29 above, and we estimate the uncertainty of $n_i$ as the counting uncertainty in a Poisson process (Eq. 3.1 in Bevington and Robinson, 2003):

$$\sigma_{n_i} = \frac{\sqrt{N_i}}{\Delta t} = \sqrt{\frac{n_i}{\Delta t}}, \quad (S31)$$

where $N_i = n_i \Delta t$ is the total particle counts during $\Delta t$.

**S1.9. Uncertainty estimation for total saltation flux by exponential profile fitting method**
When computing the total saltation flux by the exponential profile fitting method ($Q_{fit}$, Sec. 4.4.1 in main text), we calculate the associated uncertainty of total flux $\sigma_{Q_{fit}}$ through application of the error propagation formula for correlated uncertainties (Eq. S3) to the formula for total flux (Eq. 18 in main text):

$$\sigma_{Q_{fit}} = \sqrt{\left(\sigma_{q_0} z_q\right)^2 + \left(\sigma_{z_q} q_0\right)^2 + 2\sigma^2_{q_0, z_q} Q_{fit}}, \quad (S32)$$

where the exponential fit parameters and their uncertainties in Eq. S32 are obtained as in S1.3 above, and we use the value of $z_{q,LF}$ from the requisite calibration interval as an initial estimate of the saltation height uncertainty for the fitting.

**S1.10. Uncertainty estimation for total saltation flux by summation method**
We obtain the uncertainty of the total saltation flux by the summation method ($Q_{sum}$, Sec. 4.4.2 in main text) by propagating the uncertainty in $q_i$ into the individual $\Delta Q_i$ (Eq. 20 in main text), i.e.,

$$\sigma_{Q,q_i} = \sqrt{\sum_i \sigma^2_{\Delta Q_i}} = \sqrt{\sum_i \left(\Delta Q_i \frac{\sigma_{q_i}}{q_i}\right)^2}, \quad (S33)$$

where $\sigma_{Q,q_i}$ refers to the contribution of $q_i$ uncertainty to $Q$. We then also consider the contribution of $z_{q,LF}$ uncertainty to the uncertainty of $Q$:

$$\sigma_{Q,z_{q,LF}} \approx Q \frac{\sigma_{z_{q,LF}}}{z_{q,LF}}, \quad (S34)$$

where we neglect the contribution of $z_{q,LF}$ uncertainty to the individual $q_i$, as these individual uncertainties are correlated. Combining these uncertainty contributions (Eqs. S33-34) in quadrature, we have:

$$\sigma_Q = \sqrt{\sigma^2_{Q,q_i} + \sigma^2_{Q,z_{q,LF}}}. \quad (S35)$$

## S2. List of variables
Below, we list all variables described in this paper. Typical units are given in parentheses, where applicable.

**Coordinate system**
$x$, streamwise coordinate (m)
$y$, spanwise coordinate (m)
$z$, height above the bed (m)

**Generic variables for fitting and uncertainty propagation**
$x_i$, values of independent variable for fitting
$y_i$, values of dependent variable for fitting and uncertainty propagation
$\sigma_{y_i}$, uncertainties in values of dependent variable for fitting and uncertainty propagation
$y_{fit,i}$, fitted values for dependent variable
$\sigma_{y_{fit,i}}$, uncertainties in fitted values for dependent variable
$a$, generic fit or parameter value for uncertainty propagation
$\sigma_a$, uncertainty in generic fit or parameter value for uncertainty propagation
$b$, generic fit or parameter value for uncertainty propagation
$\sigma_b$, uncertainty in generic fit or parameter value for uncertainty propagation
$\sigma_{a,b}^2$ covariance in generic fit or parameter values for uncertainty propagation
$s$, generic dependent variable for uncertainty propagation
$v_i$, generic fit or parameter value for uncertainty propagation
$\sigma_{v_i}$, uncertainty in generic fit or parameter value for uncertainty propagation
$\sigma_s$, uncertainty in generic dependent variable for uncertainty propagation

**Particle variables**
$d$, particle diameter (mm)
$\bar{d}$, volume-weighted mean diameter of particles passing through the HF sensor (mm)
$d_{10}$, 10$^{th}$ percentile diameter of surface particles by volume (mm)
$\sigma_{d_{10}}$, uncertainty in 10$^{th}$ percentile diameter of surface particles by volume (mm)
$d_{50}$, median diameter of surface particles by volume (mm)
$\sigma_{d_{50}}$, uncertainty in median diameter of surface particles by volume (mm)
$d_{90}$, 90$^{th}$ percentile diameter of surface particles by volume (mm)
$\sigma_{d_{90}}$, uncertainty in 90$^{th}$ percentile diameter of surface particles by volume (mm)
$\rho_s$, particle density (kg m$^{-3}$)
$V$, particle volume (mm$^3$)

**Wind variables**
$u$, streamwise wind velocity (m s$^{-1}$)
$\bar{u}$, mean streamwise wind velocity (m s$^{-1}$)
$v$, spanwise wind velocity (m s$^{-1}$)
$\bar{v}$, mean spanwise wind velocity (m s$^{-1}$)
$w$, vertical wind velocity (m s$^{-1}$)
$\bar{w}$, mean vertical wind velocity (m s$^{-1}$)

$z_U$, anemometer height (m)
$\tau$, shear stress (Pa)
$u_*$, shear velocity (m s$^{-1}$)

**General saltation flux variables**
$q_i$, height-specific horizontal saltation flux (g m$^{-2}$ s$^{-1}$)
$\sigma_{q_i}$, uncertainty in height-specific horizontal saltation flux (g m$^{-2}$ s$^{-1}$)
$q_0$, saltation profile scaling parameter (g m$^{-2}$ s$^{-1}$)
$\sigma_{q_0}$, uncertainty in saltation profile scaling parameter (g m$^{-2}$ s$^{-1}$)
$z_q$, characteristic $e$-folding saltation layer height (m)
$\sigma_{z_q}$, uncertainty in characteristic $e$-folding saltation layer height (m)
$\sigma^2_{q_0,z_q}$, covariance of exponential flux profile fitting parameters $q_0$ and $z_q$ (g$^2$ m$^{-2}$ s$^{-2}$)
$Q$, total (vertically-integrated) saltation flux (g m$^{-1}$ s$^{-1}$)
$\sigma_Q$, total saltation flux uncertainty (g m$^{-1}$ s$^{-1}$)

**Saltation flux variables for LF traps**
$\sigma_{trap}$, relative uncertainty in mass of sand collected in LF trap (%)
$\sigma_{weight}$, absolute uncertainty in weight of sand collected in LF trap (g)
$m_{LF,i}$, mass of sand collected for $i$th LF trap (g)
$\sigma_{m_{LF,i}}$, uncertainty in mass collected for $i$th LF trap (g)
$H_{LF}$, height of LF trap opening (cm)
$\sigma_{H_{LF}}$, uncertainty in LF trap height (cm)
$T_{LF}$, duration of LF trap opening (s)
$\sigma_{T_{LF}}$, uncertainty in LF trap opening duration (s)
$W_{lF}$, width of LF trap opening (cm)
$\sigma_{W_{LF}}$, uncertainty in LF trap width (cm)
$z_{LF,i}$, height of trap opening for LF trap $i$ (m)
$\sigma_{z_{LF,i}}$, uncertainty in height of LF trap $i$ (m)
$z_{bot,LF,i}$, bottom height of trap opening for LF trap $i$ (m)
$q_{LF,i}$, height-specific horizontal saltation flux for LF trap $i$ (g m$^{-2}$ s$^{-1}$)
$\sigma_{q_{LF,i}}$, uncertainty in height-specific horizontal saltation flux for LF trap $i$ (g m$^{-2}$ s$^{-1}$)
$q_{exp,LF,i}$, exponentially fitted height-specific horizontal saltation flux for LF trap $i$ (g m$^{-2}$ s$^{-1}$)
$q_{pwr,LF,i}$, power-law fitted height-specific horizontal saltation flux for LF trap $i$ (g m$^{-2}$ s$^{-1}$)
$q_{fit,LF,i}$, generic best fit value for height-specific horizontal saltation flux for LF trap $i$ (g m$^{-2}$ s$^{-1}$)
$q_{0,LF}$, saltation profile scaling parameter for exponential profile fit to LF traps (g m$^{-2}$ s$^{-1}$)
$\sigma_{q_{0,LF}}$, uncertainty in fitted flux exponential profile scaling parameter for LF traps (g m$^{-2}$ s$^{-1}$)
$z_{q,LF}$, characteristic ($e$-folding) saltation layer height based on LF trap exponential profile fit (m)
$\sigma_{z_{q,LF}}$, uncertainty in fitted characteristic $e$-folding saltation layer height for LF trap fit (m)
$\sigma^2_{q_0,z_q,LF}$, covariance of exponential flux profile fitting parameters for LF trap fit (g$^2$ m$^{-2}$ s$^{-2}$)
$q_{p,LF}$, best-fit scaling parameter for power law fit the LF profile (variable units)
$\sigma_{q_{p,LF}}$, uncertainty in best-fit scaling parameter for power law fit the LF profile (variable units)

$k_{z,LF}$, best-fit scaling parameter and characteristic power law exponent for the LF profile (variable units)
$\sigma_{k_{z,LF}}$, uncertainty in best-fit scaling parameter and characteristic power law exponent for the LF profile (variable units)
$Q_{LF}$, estimated total (vertically-integrated) saltation flux for LF trap profile fit (g m s$^{-1}$)

**Saltation flux variables for HF sensors**
$z_{HF,i}$, height of HF sensor $i$ (g m$^{-2}$ s$^{-1}$)
$\sigma_{z_{HF,i}}$, uncertainty in height of HF sensor $i$ (g m$^{-2}$ s$^{-1}$)
$z_{dist}$, bed elevation measured by distance sensor (m)
$\sigma_{z_{dist}}$, uncertainty in bed elevation measured by distance sensor (m)
$z_{rel,HF,i}$, height of HF sensor $i$ relative to distance sensor (m)
$\sigma_{z_{rel,HF,i}}$, uncertainty in height of HF sensor $i$ relative to distance sensor (m)
$A_{HF}$, area of HF sensor opening (mm$^2$)
$n_i$, pulse count rate for HF sensor $i$ for calibration (s$^{-1}$)
$\sigma_{n_i}$, particle count rate uncertainty for HF sensor $i$ (s$^{-1}$)
$n_{cal,HF,i}$, particle counts rate for HF sensor $i$ during calibration time interval
$N_{cal,HF,i}$, total number of particles counted by HF sensor $i$ during calibration time interval
$q_{pred,HF,i}$, predicted height-specific horizontal saltation flux for HF sensor $i$ based on fitted LF trap profile (g m$^{-2}$ s$^{-1}$)
$\sigma_{q_{pred,HF,i}}$, uncertainty in predicted height-specific horizontal saltation flux for HF sensor $i$ based on fitted LF trap profile (g m$^{-2}$ s$^{-1}$)
$C_{qn,i}$, calibration factor for count rate conversion to height-specific horizontal saltation flux for HF sensor $i$ (kg m$^{-2}$)
$\sigma_{C_{qn,i}}$, uncertainty in calibration factor for HF sensor $i$ (kg m$^{-2}$)
on LF profile (g m$^{-2}$ s$^{-1}$)
$\Delta t$, duration of the sampling window
$C_{qn,pred,i}$, predicted calibration factor for HF sensor $i$ (kg m$^{-2}$)
$Q_{fit}$, total (vertically-integrated) saltation flux estimated by exponential fitting (g m$^{-1}$ s$^{-1}$)
$Q_{sum}$, total (vertically-integrated) saltation flux estimated by summation (g m$^{-1}$ s$^{-1}$)

**Supporting information: Figures**

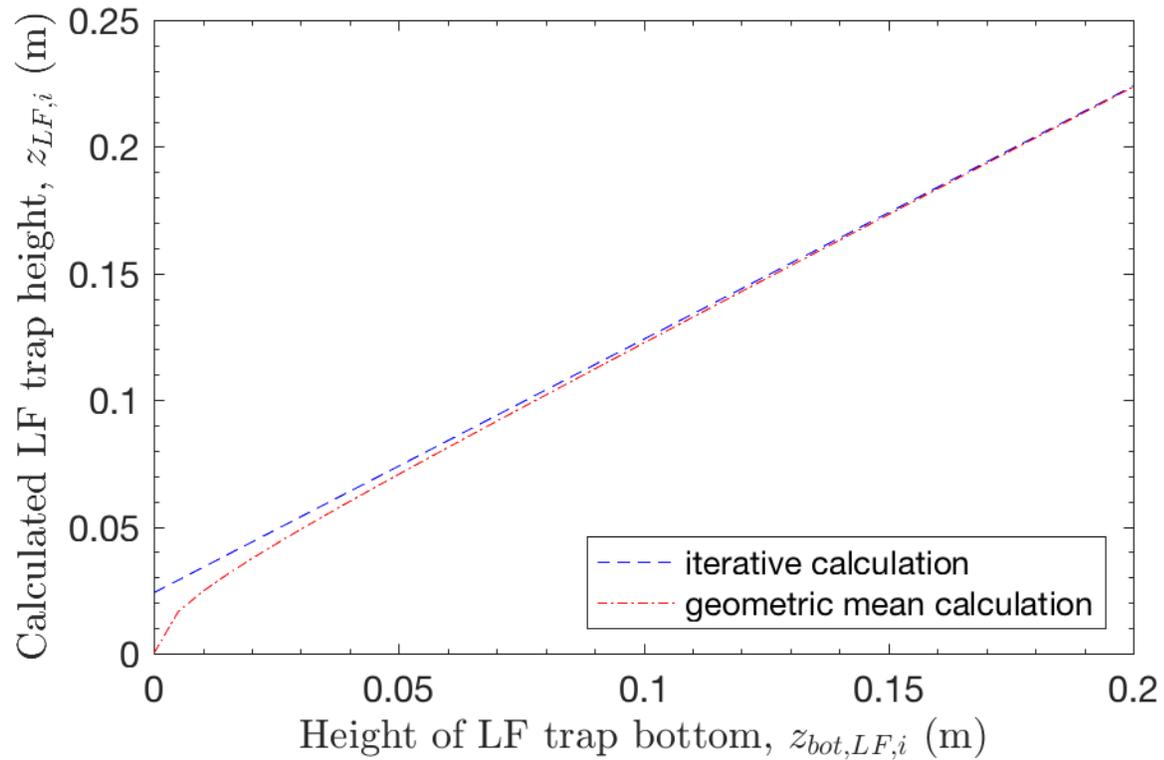

**Figure S1.** Comparison of iterative (Eq. 3) and geometric mean (Ellis et al., 2009a) methods for estimating the representative height above the bed surface for a sediment trap with $H_{LF} = 5$ cm vertical extent, assuming an exponential saltation flux profile.

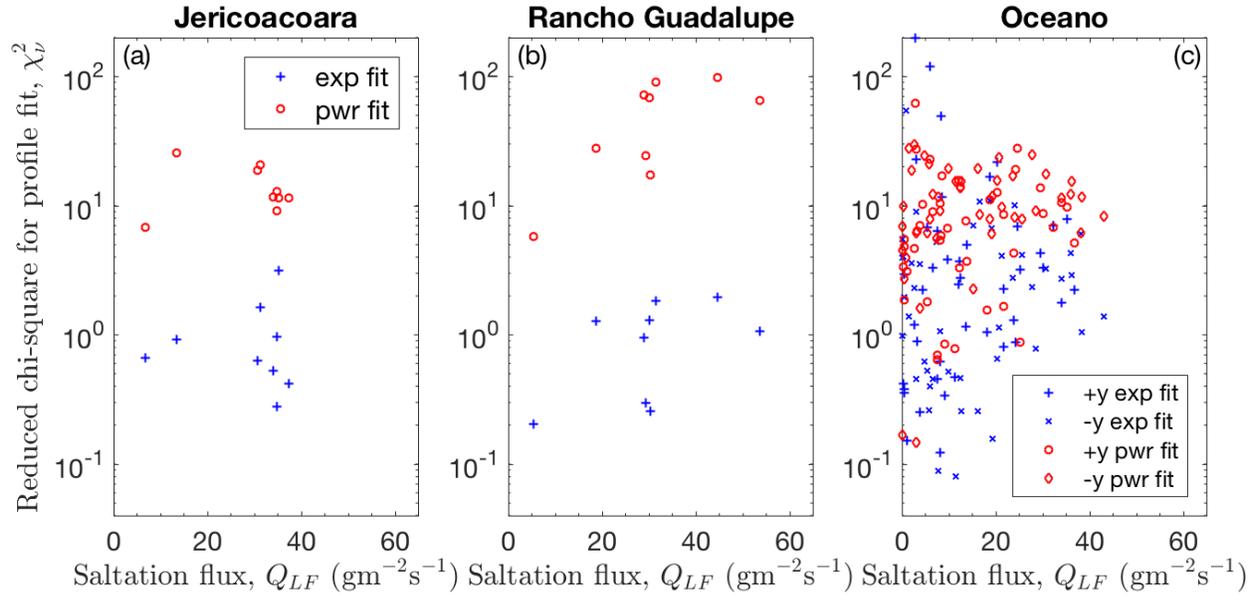

**Figure S2.** Comparison of reduced chi-square $\chi_\nu^2$ values (Eq. 7) for exponential (Eq. 3) versus power law profile fits (Eq. 5) to LF trap saltation flux profiles at the 3 field sites, compared to estimated total saltation fluxes $Q_{LF}$ for the LF traps. We calculated each $Q_{LF}$ through application of the fitting method (Eq. 18) calculated with the LF profile fit parameters $q_{0,LF}$ and $z_{q,LF}$. Smaller values of $\chi_\nu^2$ indicate a better quality of fit. At Oceano, we separated the analyses for LF traps on the right ($+y$) and left ($-y$) side of the field site, since these were subject to a spanwise separation of approximately 10 meters and therefore displayed distinctive profiles.

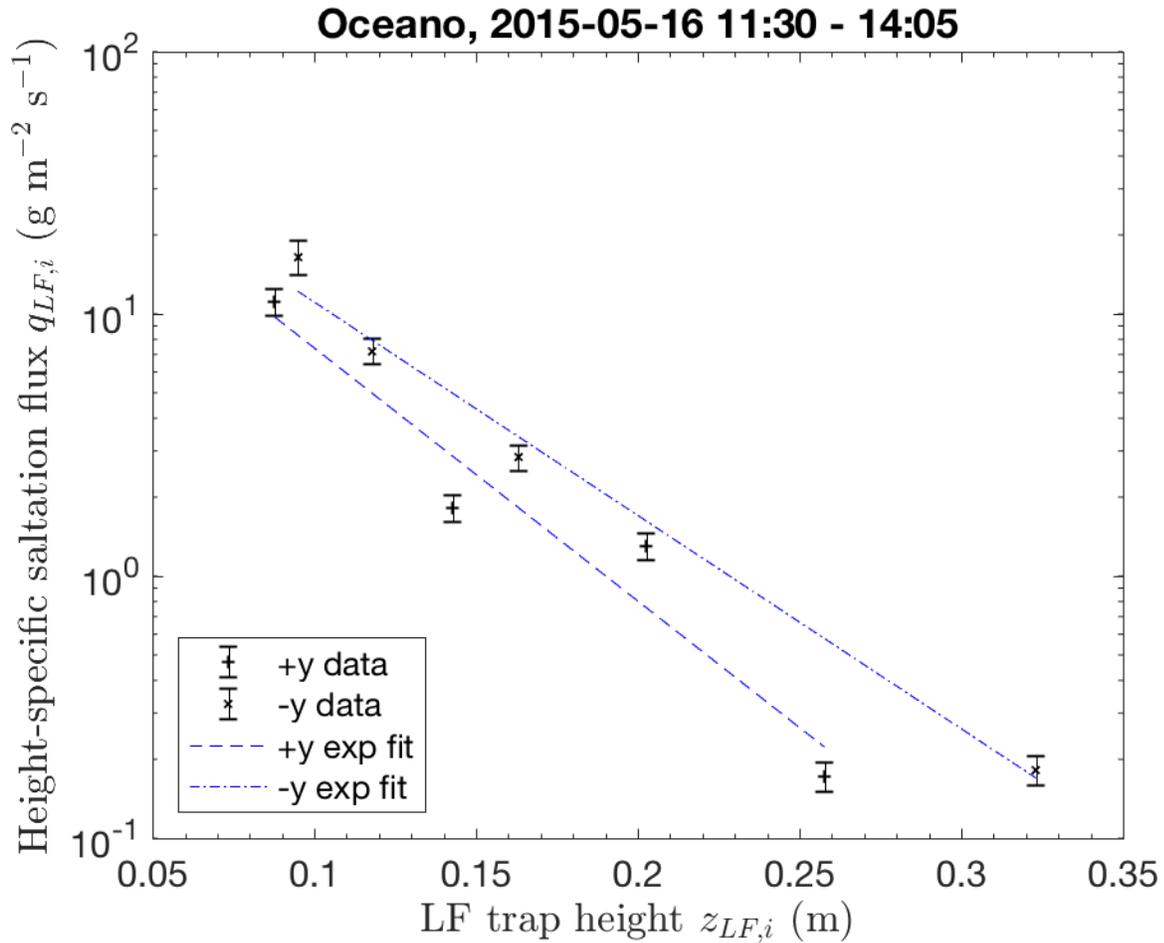

**Figure S3.** Sample LF flux profile and fits for Oceano, using BSNE traps. The data are from Oceano on 16 May 2015, from 11:30 to 14:05 local time. Due to the wide spanwise trap separation, we separately fit profiles for $+y$ and $-y$ traps, then averaged the best fit parameters from the two profiles (Eqs. 8-9).

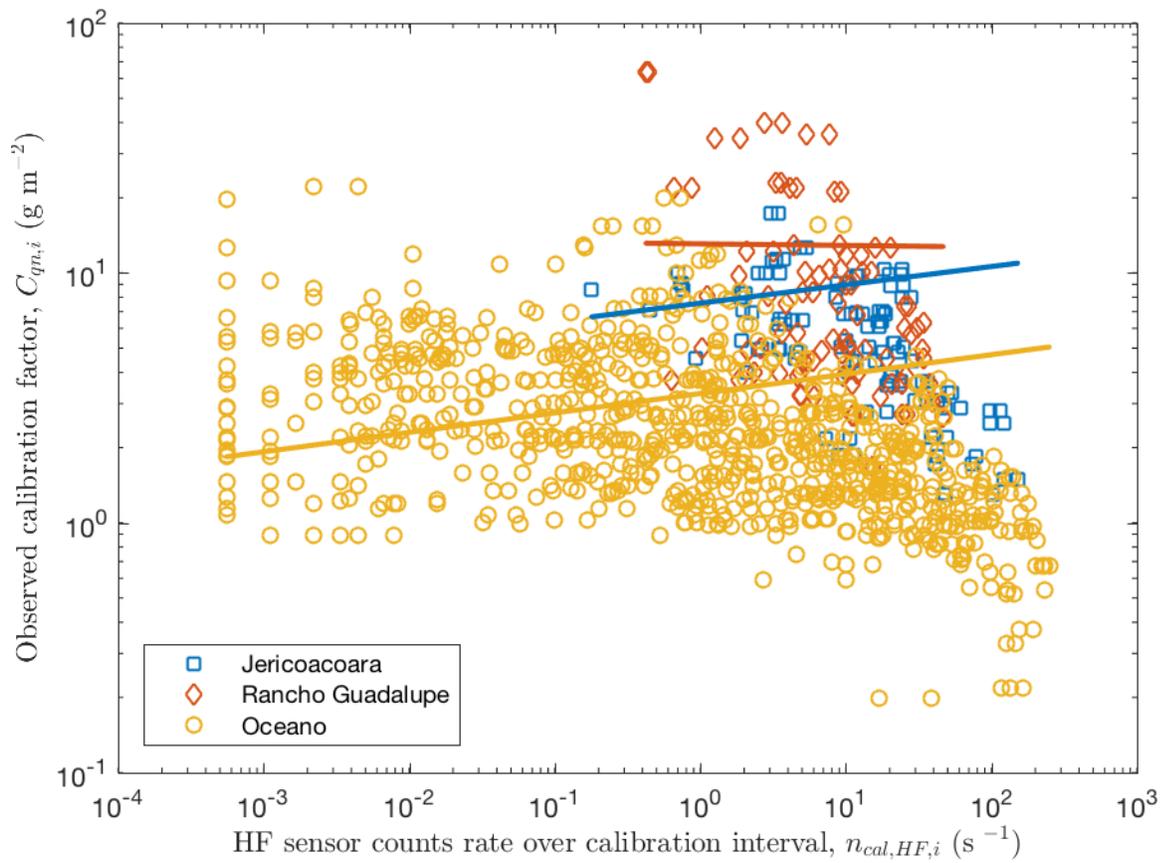

**Figure S4.** Variation of calibration factor $C_{qn,i}$ with HF sensor particle counts rate over calibration interval $n_{cal,HF,i}$. Solid lines show linear fit to $\log(C_{qn,i})$ versus $\log(n_{cal,HF,i})$, indicating slight but statistically significant increase in $C_{qn,i}$ with $n_{cal,HF,i}$ at Jericoacoara and Oceano.